\documentclass[onecolumn,floatfix,superscriptaddress,a4paper,
               showpacs,showkeys,nofootinbib,preprint]{revtex4}
\usepackage{epsfig}
\usepackage{latexsym}
\usepackage{xspace}
\usepackage{hyperref}
\usepackage[latin2]{inputenc}
\usepackage{indentfirst}
\usepackage{enumerate}

\usepackage{amsmath}
\usepackage{amssymb}
\usepackage[english]{babel}
\usepackage{url}

\newcommand{\eq}[1]{\begin{align} #1 \end{align}}

\newcommand*{\dis}{\displaystyle}

\begin{document}

\title{System-size and energy dependence
of particle momentum spectra:  \\ The UrQMD analysis of p+p and Pb+Pb collisions}
\author{V. Yu. Vovchenko}
\affiliation{
Taras Shevchenko National University of Kiev, Kiev, Ukraine}
\affiliation{
Frankfurt Institute for Advanced Studies, Frankfurt, Germany}
\affiliation{
Goethe University, Frankfurt, Germany}
\author{D. V. Anchishkin}
\affiliation{
Bogolyubov Institute for Theoretical Physics, Kiev, Ukraine}
\affiliation{
Taras Shevchenko National University of Kiev, Kiev, Ukraine}
\author{M. I. Gorenstein}
\affiliation{
Bogolyubov Institute for Theoretical Physics, Kiev, Ukraine}
\affiliation{
Frankfurt Institute for Advanced Studies, Frankfurt, Germany}

\date{\today}


\begin{abstract}
The UrQMD transport model is used to study a system-size and energy dependence
of the pion production in high energy collisions.
New data of  the NA61/SHINE Collaboration  on
spectra of negatively charged pions in proton-proton  interactions at SPS energies
are considered. These results are compared with the corresponding data of the NA49 Collaboration
in central Pb+Pb collisions at the same collision energies per nucleon.
Mean pion multiplicity per participant nucleon, inverse slope parameter of the
transverse momentum spectra, and width of rapidity distribution are
investigated. A role of isospin effects is  discussed.
We find that the UrQMD  model predicts a non-monotonous behavior of
mean transverse mass with collision energy for positively charged pions at the
mid-rapidity in inelastic proton-proton  interactions.
This will be checked soon experimentally by NA61/SHINE Collaboration.
\end{abstract}

\maketitle

\section{Introduction}

During the period of 1999-2002, experimental data on
hadron production in Pb+Pb collisions at $p_{\rm lab}=$20$A$, 30$A$, 40$A$, 80$A$, and 158$A$~GeV/c   were recorded by the NA49 Collaboration at the Super Proton Synchrotron (SPS) of the European Organization for Nuclear
Research (CERN) \cite{NA49-1,NA49-3,NA49-2}.
These results  are consistent with the onset of
deconfinement in central Pb+Pb collisions at about 30$A$~GeV/c \cite{GaGo}.
A further progress in understanding the effects related to the onset of
deconfinement can be achieved by a new comprehensive study of hadron production
in proton-proton, proton-nucleus, and nucleus-nucleus collisions.
This  motivated the present NA61/SHINE ion programme at the SPS CERN devoted to
the system size and energy scan~\cite{Ga:2009,NA61facility}.
In parallel to the NA61/SHINE programme the Beam
Energy Scan (BES) Programme at the BNL RHIC was proposed~\cite{RHIC}.
Both programmes
were motivated by the NA49 results on the onset
of deconfinement and the possibility to observe the critical
point  of strongly interacting matter.
These efforts will be extended by the future
Compressed Baryonic Matter (CBM) experiment at
the Facility for Antiproton and Ion Research (FAIR)~\cite{CBMPhysicsBook}.
It employs high  intensity beams and large acceptance detectors,
and will make it possible to measure rare probes~--
multi-strange hyperons and open charmed  mesons -- in proton-proton, proton-nucleus,
and nucleus-nucleus collisions.
The CBM experiment~\cite{cbm}
is expected to start
data taking with the beams up to 11$A$~GeV from
SIS100 in 2018.
A considered addition of SIS300 would  increase the energy range to 35$A$~GeV
after 2025.

Recently, the NA61/SHINE Collaboration published data \cite{NA61pp}
on spectra of negatively charged pions  produced in inelastic proton-proton (p+p)
interactions at $p_{\rm lab}=$~20, 31, 40, 80, and 158~GeV/c.
This is the first experimental step of the system-size scan of NA61/SHINE.
Data on $^7$Be+$^9$Be collisions are already recorded, collisions of Ar+Ca and Xe+La
are planned to be registered in 2015 and 2017, respectively.

In the present study the spectra of negatively charged pions in inelastic p+p collisions
are considered, and a comparison with the corresponding data on central Pb+Pb
collisions is performed.
We analyze changes between p+p and Pb+Pb collisions with regards
to different hadron observables found in experimental data and
compare them to those predicted by
theoretical modeling, which
is based on the ultra-relativistic quantum
molecular dynamics (UrQMD-3.3p2) model~\cite{UrQMD1998,UrQMD1999}.
In the present analysis we use the cascade version of UrQMD without  an
intermediate hydro stage.
The energy dependence of main characteristics of pion spectra are
investigated. 
It is shown that a proper treatment of isospin effects is important for
comparison of pion production data in proton-proton and nucleus-nucleus
collisions.

The paper is organized as follows.
In Sec.~\ref{iso-spin} pion multiplicities per participant nucleon are
considered.
The isospin effects are also discussed. In Sec.~{\ref{sec-loss} the UrQMD model
is used to calculate
energy loss of colliding nucleons in inelastic
p+p and central nucleus-nucleus collisions.
In Sec. \ref{sec-spec} spectra of $\pi^-$ mesons produced in inelastic p+p and
central Pb+Pb collisions are compared and analyzed with the UrQMD
model.
A summary in Sec.~\ref{sum} closes the article.

\section{Pion Multiplicities and Isospin Effects}
\label{iso-spin}

Mean multiplicity of pions per participant nucleon as well as
properly normalized pion spectra are quite similar in inelastic
proton-proton and nucleus-nucleus collisions at the same collision energy per nucleon  in the SPS energy range.
Therefore, one needs comprehensive  data at the same collision energies per
nucleon to make a systematic comparison of the pion spectra in p+p and
nucleus-nucleus collisions and reveal the physical differences between
them.
The recent NA61/SHINE data on p+p interactions \cite{NA61pp} and the data of NA49 on Pb+Pb
collisions \cite{NA49-1,NA49-3,NA49-2} will be used in the present analysis.

Similar behavior of the pion spectra in proton-proton and nucleus-nucleus
collisions explains also a particular vitality of the wounded
nucleon model (WNM) \cite{WNM} which treats the final state in nucleus-nucleus
collision as the result of independent
particle production from wounded (participant) nucleons.
Meanwhile,
the WNM model failed to explain the production of
strange hadrons and anti-baryons: for these secondary particles a strong
difference is observed between proton-proton and nucleus-nucleus collisions.

The UrQMD model will be used for our analysis with event statistics of about
$2\cdot 10^6$ inelastic p+p reactions and $5\cdot 10^4$ central Pb+Pb collisions.
For Pb+Pb collisions the experimental centrality selection is $7.2\%$
at $p_{\rm lab}=$20$A$, 30$A$, 40$A$, 80$A$~GeV/c and $5\%$ at $p_{\rm lab}=158A$~GeV/c.
In the UrQMD simulations these samplings correspond
to the restrictions on the impact parameter $b<4$~fm and $b<3.4$~fm, respectively.
In fixed target experiments, the centrality selection is made using
the number of projectile spectators measured by  a zero degree calorimeter.
However,
for the first moments, such as the mean number of some particle species,
any centrality criteria
appears to be equivalent to the geometrical one treated with a help of
the impact parameter $b$ \cite{BF}.
This is not valid for the studies of event-by-event fluctuations (see, e.g.,
Ref.~\cite{fluc}).

The experimental results for mean number of negatively charged pions
$\langle \pi^-\rangle$ divided by average number of participant nucleons
$\langle N_p\rangle$ (equal to two for p+p collisions)
are shown in Fig.~\ref{fig:NoverNp}.
In this figure the energy measure
$F = \left(\sqrt{s_{_{NN}}}-2m_N\right)^{3/4}/\left(\sqrt{s_{_{NN}}}\right)^{1/4}$ is used,
where $\sqrt{s_{_{NN}}}$ is the center of mass energy of the nucleon pair and
$m_N$ is the nucleon mass. The laboratory energy
per nucleon is equal to $E_{\rm lab}/{\rm A}=(s_{NN}-2m_N^2)/2m_N$.
From Fig.~\ref{fig:NoverNp} it is seen that the quantity
$\langle \pi^-\rangle/\langle N_p\rangle$ in central Pb+Pb
collisions is larger than in inelastic p+p collisions at all SPS energies.
The lines correspond to the UrQMD results.
The UrQMD result for inelastic p+p interactions shows a good agreement with the
data
for the top SPS energies
(80 and 158~GeV) but underestimates the experimental values of
$\langle \pi^-\rangle/\langle N_p\rangle$
at the lower ones.
\begin{figure}
\begin{center}
\begin{minipage}{.65\textwidth}
\centering
\includegraphics[width=\textwidth]{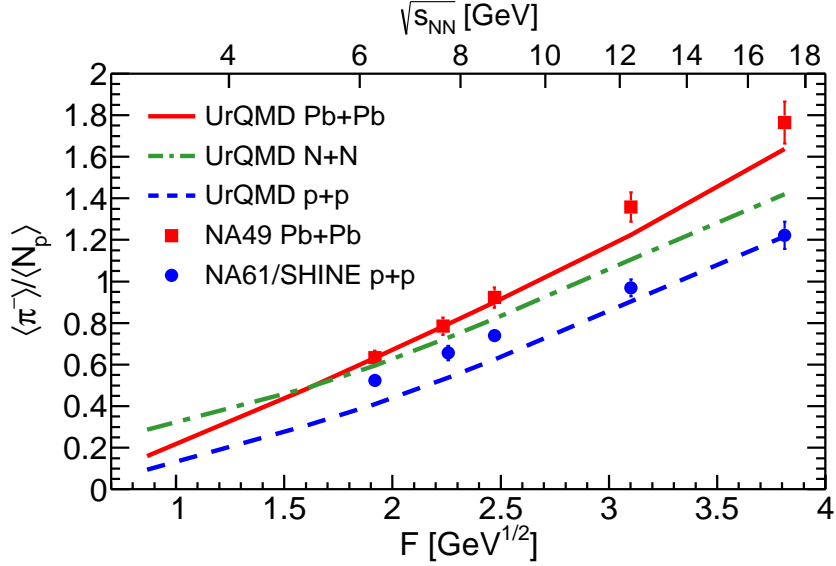}
\end{minipage}
\caption{(Color online) Energy dependence of mean $\pi^-$ multiplicity divided by
the average number of
participant nucleons. The data of NA49 on central
Pb+Pb collisions and NA61/SHINE on inelastic p+p interactions are depicted by symbols.
The lines correspond to the
UrQMD predictions for inelastic p+p (dashed),
central Pb+Pb (solid),
and inelastic N+N (dashed-dotted)
collisions.
}
\label{fig:NoverNp}
\end{center}
\end{figure}

\begin{figure}
\begin{minipage}{.48\textwidth}
\centering
\includegraphics[width=\textwidth]{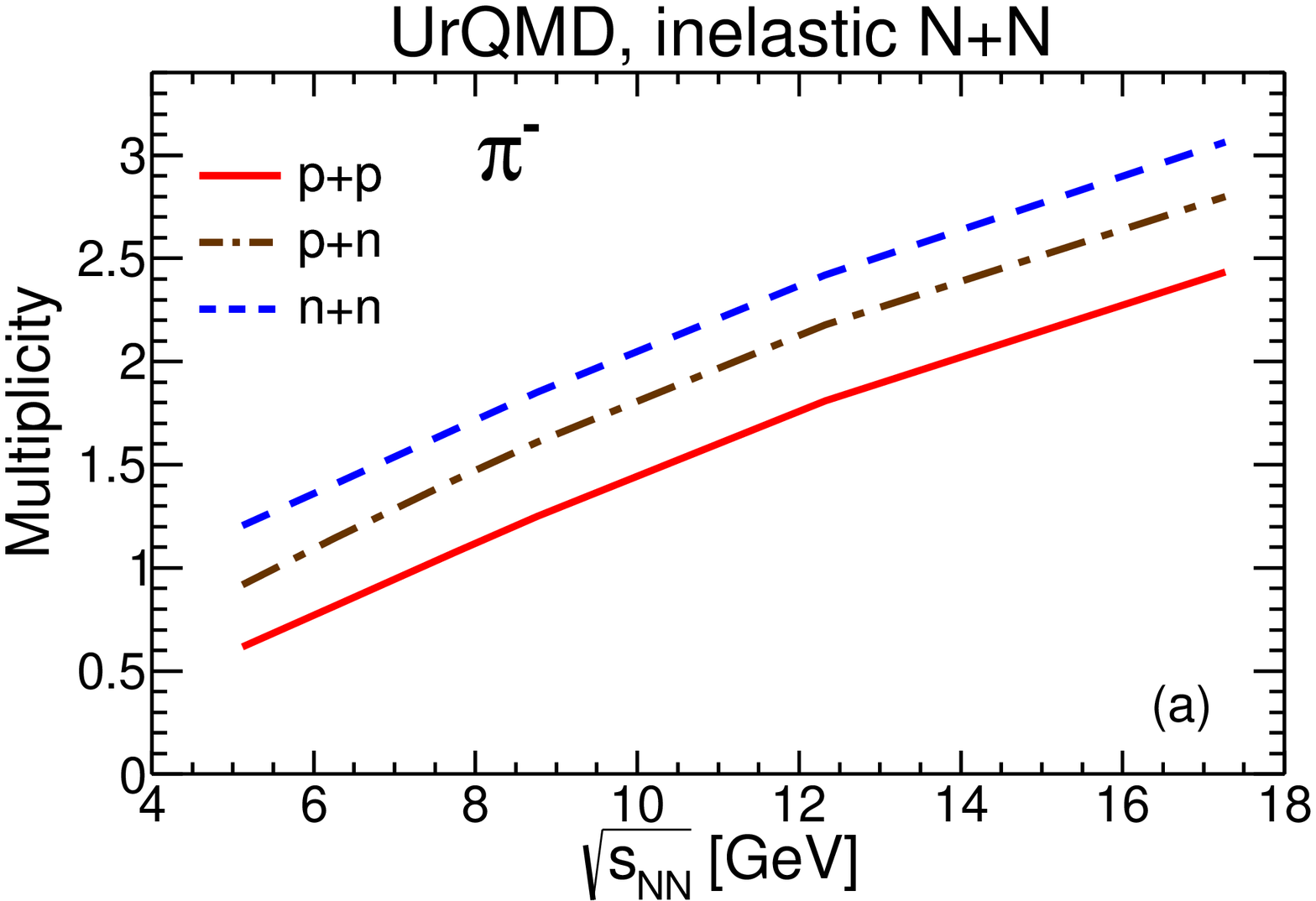}
\end{minipage}
\begin{minipage}{.48\textwidth}
\centering
\includegraphics[width=\textwidth]{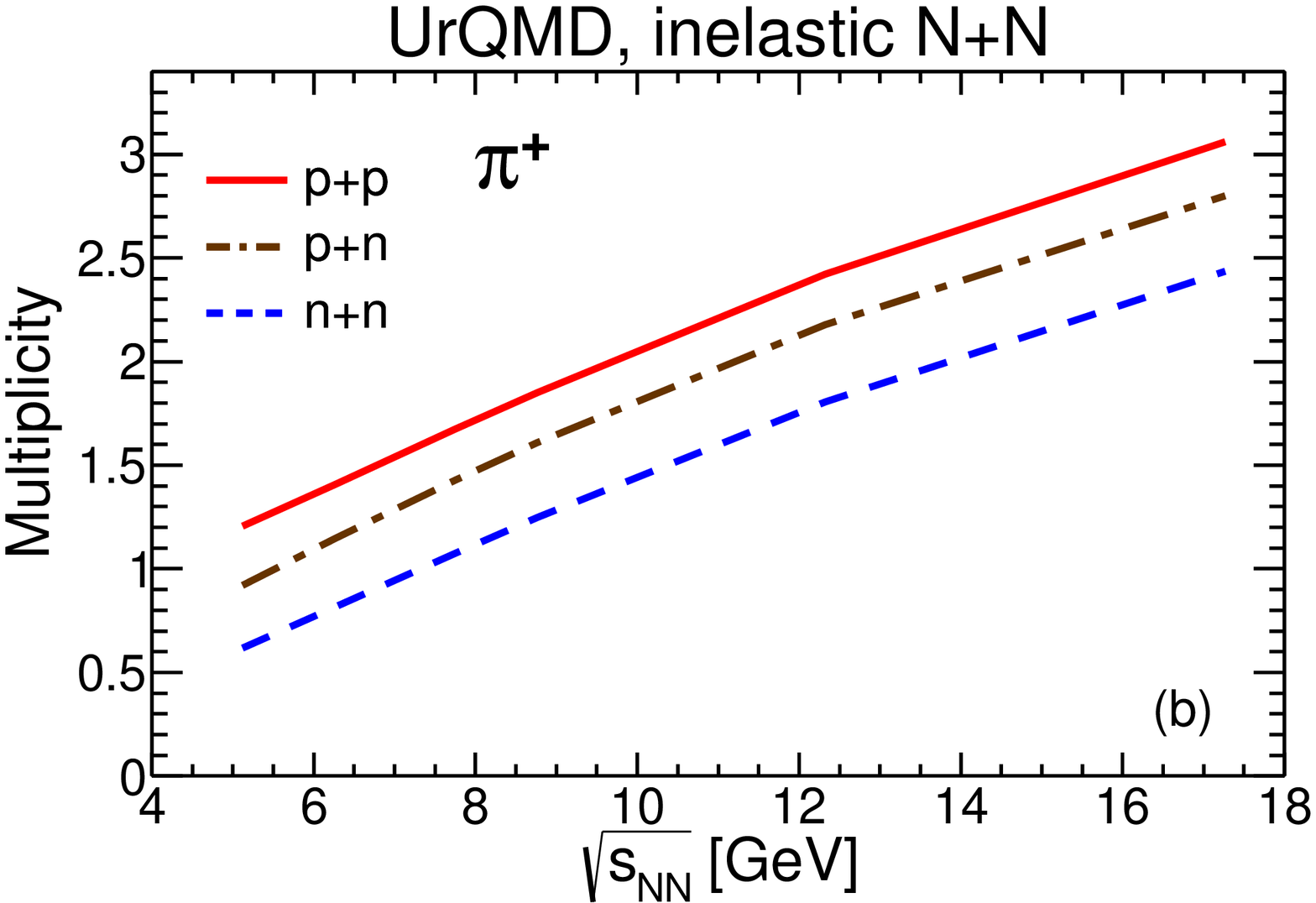}
\end{minipage}
\caption{(Color online) Collision energy dependence of mean multiplicity of (a) $\pi^-$ and
(b) $\pi^+$ mesons in inelastic p+p, p+n and n+n interactions at SPS energies calculated
in the UrQMD model.}
\label{fig:multinelNN}
\end{figure}

To make an accurate comparison of mean pion multiplicities and some other
quantities in proton-proton and nucleus-nucleus collisions, one needs to take
into account the isospin effects.
In nucleus-nucleus reaction, different types of nucleon-nucleon (NN) collisions
take place:  p+p, p+n, and n+n.
The $\pi^-$ and $\pi^+$ multiplicities in inelastic p+p, p+n, and n+n collisions
calculated at the SPS energies within the UrQMD model are presented  in
Fig.~\ref{fig:multinelNN}.
The differences between the $\pi^-$ or $\pi^+$ multiplicities in different
types of N+N collisions are approximately constant.
Thus, the largest relative isospin effects are evidently expected at smallest
energies where the absolute values of the pion multiplicities are small.

Let us consider A+A collision as a set of  p+p,
proton-neutron (p+n),
and neutron-neutron (n+n) reactions.  Their partial contributions 
to hadron production in A+A collision will
depend on the atomic number $A$ and electric charge $Z$
of colliding nuclei.      
Hadron multiplicities in  N+N reaction will be defined as an
appropriate average over all possible p+p, p+n, and n+n collisions taking place
in A+A reaction, e.g., for $\pi^i$ multiplicity ($i=+,-,0$) one
obtains:
\begin{equation}\label{NN}
\langle \pi^i_{NN} \rangle~ = ~\alpha_{pp} \ \langle \pi^i_{pp} \rangle
~+~ \alpha_{pn} \ \langle \pi^i_{pn} \rangle ~ +~ \alpha_{nn} \
\langle \pi^i_{nn} \rangle\,,
\end{equation}
where $\alpha_{pp}=Z^2/A^2$, $\alpha_{pn}=2(A-Z)Z/A^2$, and
$\alpha_{nn}=(A-Z)^2/A^2$ are the combinatoric probabilities
of p+p, p+n, and n+n collisions, respectively. 
For light and intermediate nuclei, $Z\cong 0.5 A$, whereas
for heavy ones, $Z\cong 0.4 A$. Therefore, $\alpha_{pp}\cong 0.16$, 
$\alpha_{pn}\cong 0.48$, and
$\alpha_{nn}\cong 0.36$ for Pb+Pb collisions, and $\alpha_{pp}\cong \alpha_{nn}\cong 0.25$ and
$\alpha_{nn}\cong 0.5$ in collisions of light and intermediate nuclei. It is clear
that relative importance of n+n reactions in A+A collisions increases with atomic number.
To avoid a misunderstanding let us stress again that the notion of N+N reaction in Eq.~(\ref{NN})
leads to different N+N results for different colliding nuclei.

The behavior of $\langle \pi^-_{NN} \rangle/2$,
which is a mean multiplicity of $\pi^-$ mesons per participant nucleon in a fictitious N+N collision,
calculated in UrQMD
with Eq.~(\ref{NN})
is shown in Fig.~\ref{fig:NoverNp} by the dashed-dotted line.
From this figure, one concludes that
an essential part of the difference in $\langle \pi^-\rangle/\langle N_p\rangle$
for p+p and Pb+Pb collisions comes from the isospin effects.

\begin{figure}
\begin{center}
\begin{minipage}{.48\textwidth}
\centering
\includegraphics[width=\textwidth]{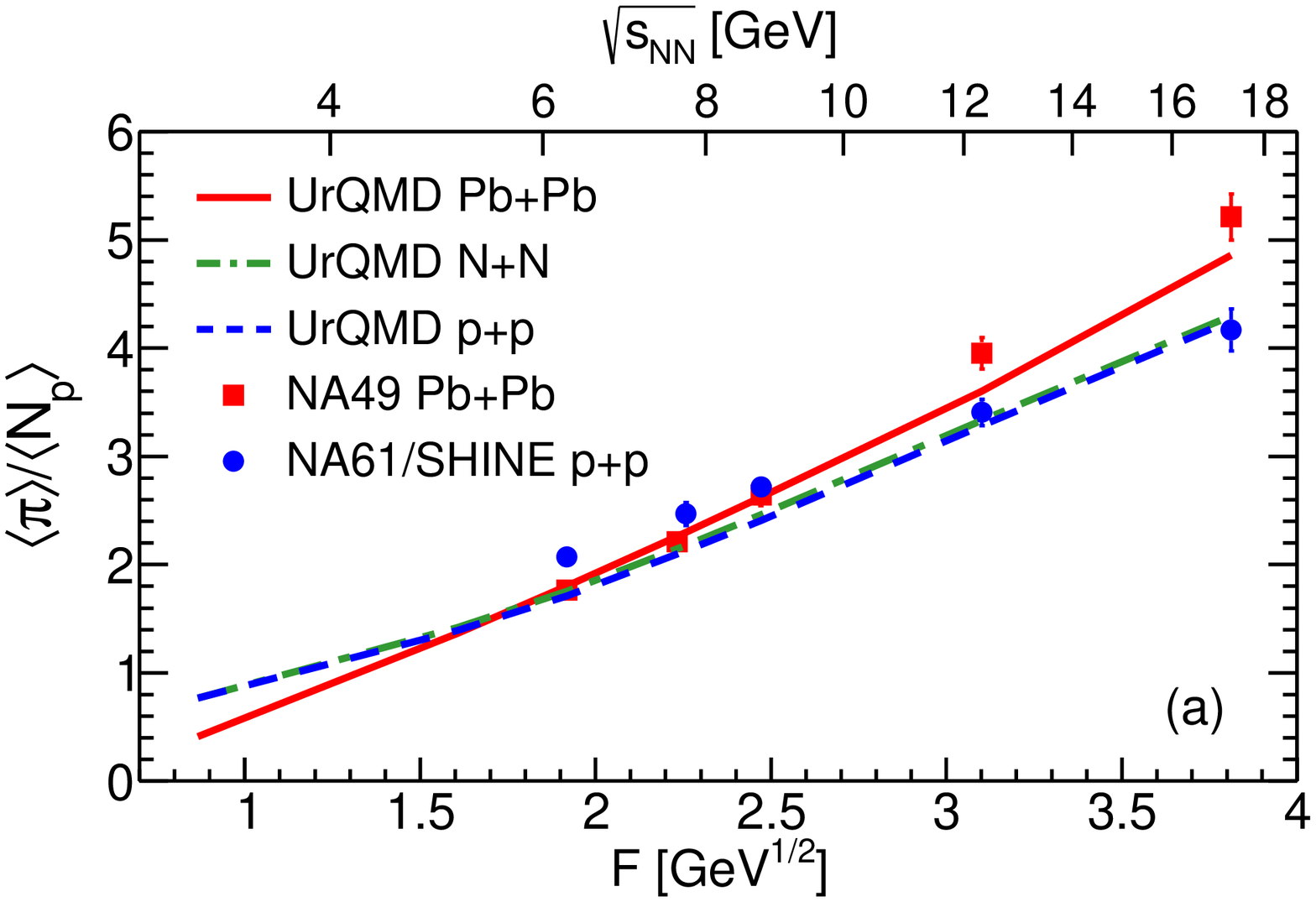}
\end{minipage}
\begin{minipage}{.48\textwidth}
\centering
\includegraphics[width=\textwidth]{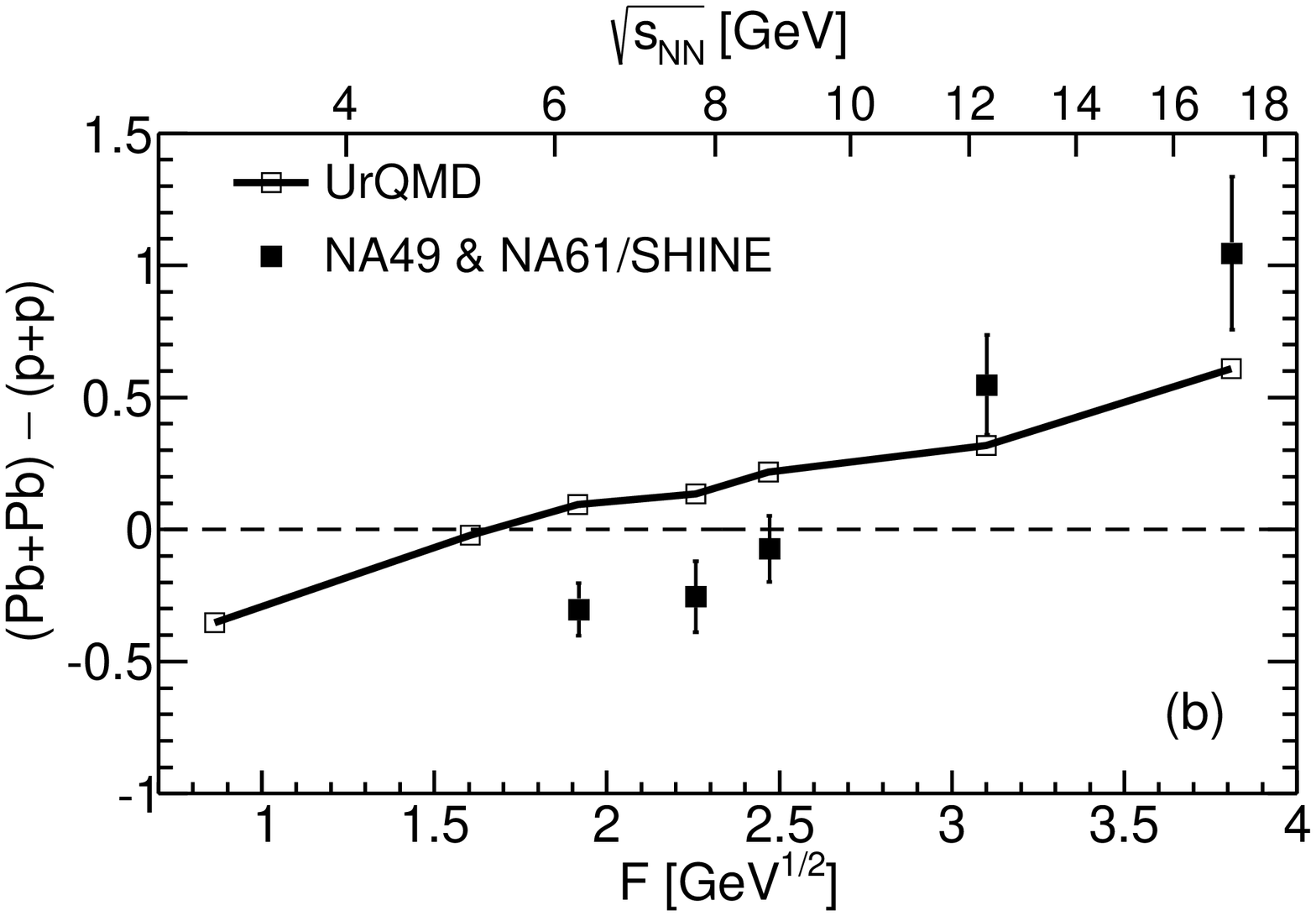}
\end{minipage}
\caption{(Color online) (a) The same as in Fig.~\ref{fig:NoverNp} but
for mean multiplicity of all pions.
(b) The difference of the mean pion multiplicity per nucleon
participant in central Pb+Pb and inelastic p+p collisions.
The full symbols correspond to the data of  NA49 and NA61/SHINE collaborations.
The open symbols  connected by solid line show the UrQMD results.  }
\label{fig:Npi}
\end{center}
\end{figure}
The results of UrQMD calculations for
mean multiplicity of all pions divided by the mean number
of participant nucleons are shown in Fig.~\ref{fig:Npi}a.
These results are compared to experimental data from Ref.~\cite{NA61pp}.
The mean multiplicity of all pions
is calculated in UrQMD directly as $\langle \pi \rangle =
\langle \pi^+ \rangle + \langle \pi^0 \rangle + \langle \pi^- \rangle$.
The experimental estimate of the mean pion multiplicity
is calculated in accordance with the formula
\begin{equation}
\langle \pi\rangle ~=~\frac 32 \ \Big(\langle \pi^+
\rangle~+~\langle \pi^-\rangle\Big)~.
\label{pi}
\end{equation}
It assumes $\langle \pi^0\rangle =0.5 \,(\langle \pi^+\rangle +\langle \pi^-\rangle)$ for
the mean multiplicity of neutral pions. Both $\langle \pi^+\rangle$ and $\langle \pi^-\rangle$
were measured in Pb+Pb central collisions. However, no systematic data for $\langle \pi^+\rangle$
in p+p reactions are now available. Thus, to construct experimental estimate of $\langle \pi\rangle$
in p+p reactions,  Eq.~(\ref{pi})
has been supplemented in Ref.~\cite{NA61pp} by an approximate relation
$\langle \pi^+\rangle\cong \langle \pi^-\rangle +2/3$ \cite{golokh}. This relation appears
to be also approximately valid for the UrQMD results. However, we do not need it in our
UrQMD modeling.  
The UrQMD results
for $\langle \pi^+\rangle $, $\langle \pi^0 \rangle$, $\langle \pi^-\rangle$ in p+p, p+n, and n+n reactions
are used to construct $\langle \pi\rangle$
in p+p and N+N collisions. Both UrQMD values for the mean pion multiplicity  -- in p+p
and N+N reactions -- are shown in Fig~\ref{fig:Npi}.  

The mean pion multiplicity per participant nucleon is larger in inelastic p+p
interactions than that in central Pb+Pb collisions at small SPS energies.
A possible explanation of the pion suppression in A+A collisions
at small energies were discussed in Ref.~\cite{GGM}, where the system 
of pions, nucleons, and deltas  were considered. When the system created
in A+A collisions evolves towards chemical equilibrium, the initial surplus 
of deltas created at the early stage of A+A reaction in p+p, n+n, and n+n collisions
has to be reduced. The microscopic process responsible for the $\Delta$ absorption
is $\Delta+N\rightarrow N+N$ and is included in UrQMD. This causes the suppression of the final state pions.    
 
The difference between the $\langle \pi\rangle/\langle N_p\rangle$ 
data values in Pb+Pb and p+p collisions increases with collision energy and
changes sign (becomes positive) at about $E_{lab}\approx 40A$~GeV
as shown in Fig.~\ref{fig:Npi}b.
This {\it kink}   in the difference of pion multiplicities per participant nucleon
in Pb+Pb and p+p collisions
was predicted by the Statistical Model of the Early
Stage \cite{GaGo} as the consequence of the onset of deconfinement.
As seen from Fig.~\ref{fig:Npi} the UrQMD results also show this type of behavior.
However, the difference
changes sign at significantly lower energy of about $E_{lab}\approx 10A$~GeV instead of $40A$~GeV.

\section{Energy Loss of Colliding Nucleons}
\label{sec-loss}

We introduce the quantity $K_{\rm loss}$ which shows a fraction of the initial  kinetic
energy lost by each nucleon  from the colliding systems to create new secondary
particles.
In the center of mass system of colliding protons or nucleus this energy loss
is defined as
\begin{equation}
K_{\rm loss} ~=~ \frac{\sqrt{s_{_{NN}}}\,/2-(\langle E_{B}\rangle ~-~\langle E_{\overline{B}} \rangle)/\langle N_p\rangle}{\sqrt{s_{_{NN}}}\,/2-m_N}
\label{eq:Kloss}
\end{equation}
where $\langle E_{B}\rangle$ and  $\langle E_{\overline{B}}\rangle $ are the
average values of the energies of baryons and anti-baryons, respectively,
in the final state (the nucleon spectators in nucleus-nucleus
collisions are excluded) and $\sqrt{s_{_{NN}}}\, /2$ is the center of mass
energy of a nucleon in the colliding nuclei.

The momentum spectra of baryons and anti-baryons were not yet presented in p+p
collisions by NA61. Therefore, one can only rely
on the model estimates.
The quantity $K_{\rm loss}$ calculated within the UrQMD simulations for inelastic p+p and central Pb+Pb collisions is shown
in Fig.~\ref{fig:Kloss}a.
It is seen that, for Pb+Pb collisions, $K_{\rm loss}$
increases with collision energy. This is consistent
with data of NA49 on
inelasticity $K$~\cite{StrobeleNA49}, which is a very closely related quantity and is depicted by triangular symbols in Fig.~\ref{fig:Kloss}a.
The UrQMD results demonstrate that at  $E_{\rm lab}\geq20$A~GeV the
following inequality holds
\begin{equation}
K_{\rm loss}({\rm p+p})~ <~ K_{\rm loss}({\rm Pb+Pb})~,
\label{loss}
\end{equation}
i.e., each nucleon in Pb+Pb collisions transforms more initial kinetic
energy to the production of new particles than that in p+p inelastic
collisions  at the same initial energy per participant nucleon,
$\sqrt{s_{_{NN}}}\, /2$.
As seen from Fig.~\ref{fig:Npi}, the relation (\ref{loss}) is correlated with
the inequality for the total pion multiplicity per participant nucleon
\begin{equation}
\left(\frac{\langle \pi\rangle}{\langle N_p\rangle} \right)_{p+p}~<~
\left(\frac{\langle \pi\rangle}{\langle N_p\rangle} \right)_{Pb+Pb}~.
\label{NpiNp}
\end{equation}
\begin{figure}
\begin{center}
\begin{minipage}{.49\textwidth}
\centering
\includegraphics[width=\textwidth]{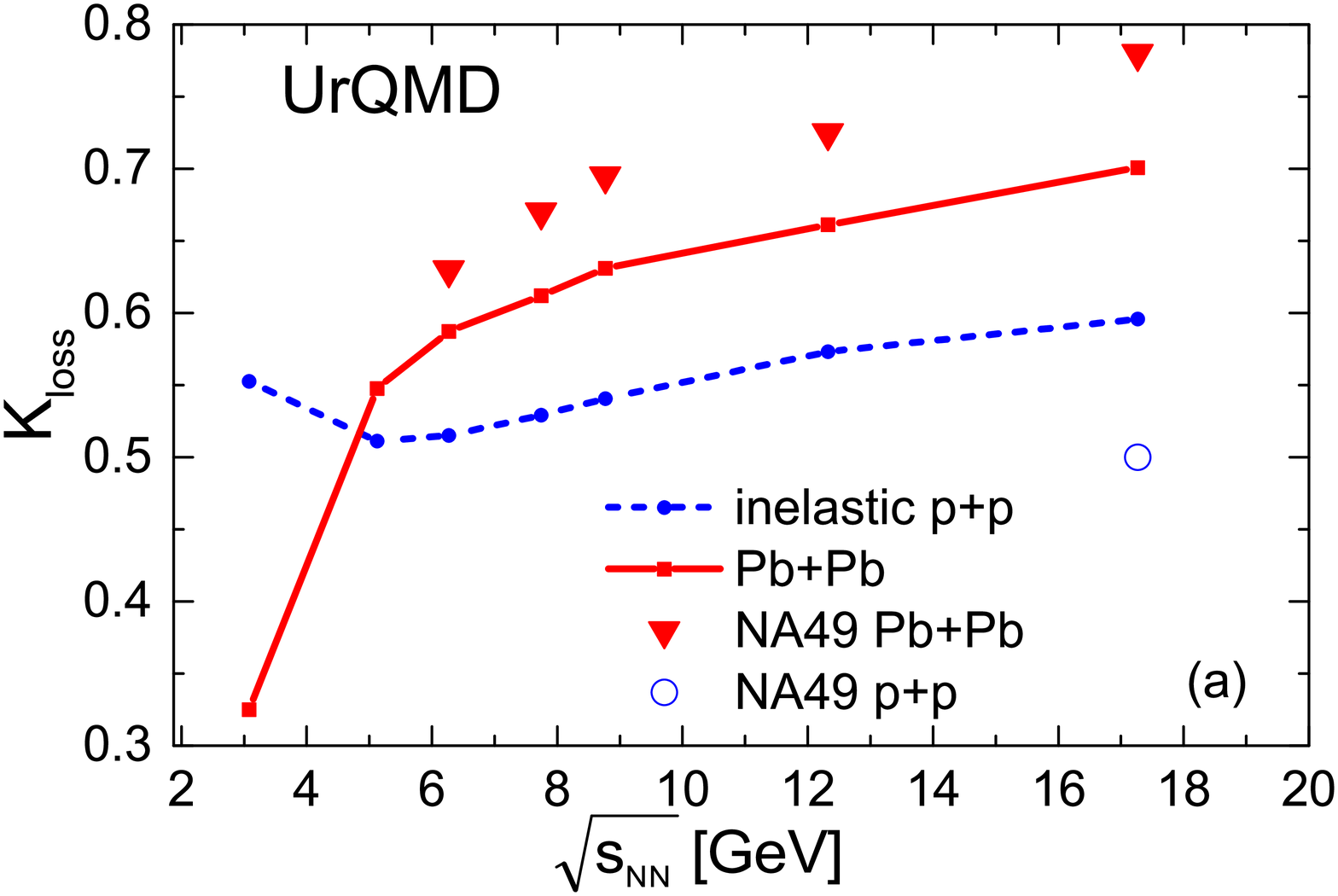}
\end{minipage}
\begin{minipage}{.49\textwidth}
\centering
\includegraphics[width=\textwidth]{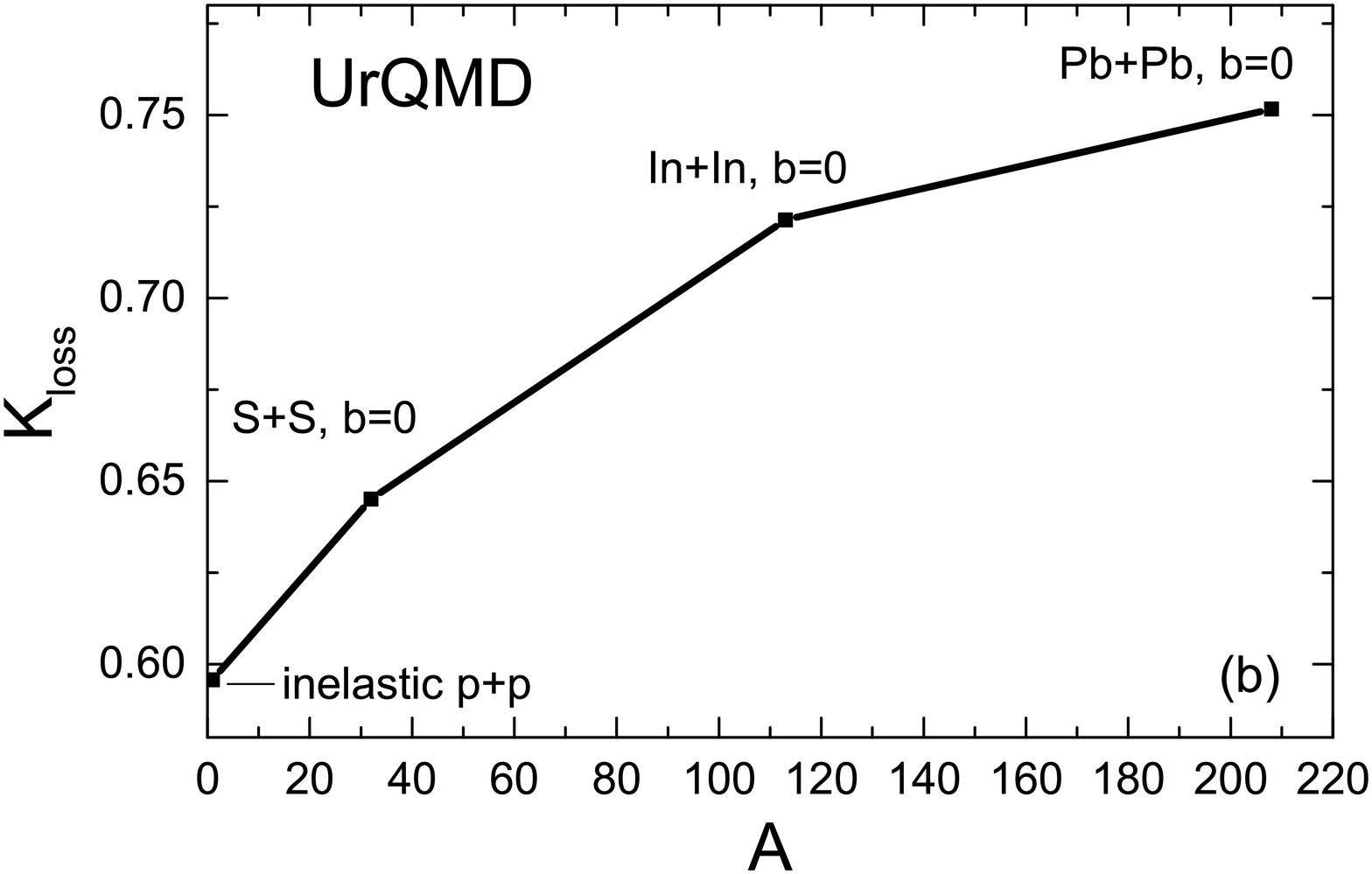}
\end{minipage}
\caption{(Color online) The UrQMD results for the quantity $K_{\rm loss}$ (\ref{eq:Kloss}).
(a) The energy dependence in inelastic p+p (the dashed line)
and Pb+Pb (the solid line) collisions.
Triangular symbols depict the Pb+Pb data and a circle corresponds
to the p+p data presented by the NA49 collaboration in Ref.~\cite{StrobeleNA49}
for the inelasticity. (b) The A-dependence
for most central (impact parameter $b=0$) nucleus-nucleus collisions at fixed collision energy of 158$A$~GeV.}
\label{fig:Kloss}
\end{center}
\end{figure}

To study the dependence of energy loss (\ref{eq:Kloss}) on the system size we
calculate its UrQMD values in central (impact parameter $b=0$) nucleus-nucleus
collisions at different mass number A of the colliding nuclei: S+S (A=32)
and In+In (A=113) collisions at $p_{\rm lab}=158A$~GeV/c.
Results of the calculations shown in Fig.~\ref{fig:Kloss}b
correspond to monotonous increase of $K_{\rm loss}$  with the mass number
of colliding nuclei at high SPS energies.
This increase of $K_{\rm loss}$ can be attributed to an increased number of participant nucleons:
a probability for secondary inelastic collisions (and thus for the
additional energy loss) of participant nucleons evidently
increases with $N_p$.

On the other hand, Figs.~\ref{fig:Kloss} and \ref{fig:Npi} show that at small
collision energies $E_{\rm lab}\le 10$~GeV the inequalities (\ref{loss}) and
(\ref{NpiNp}) found from the UrQMD simulations are changed to opposite ones:
\begin{eqnarray}
\label{loss-small}
& K_{\rm loss}(p+p)~>~ K_{\rm loss}(Pb+Pb)~,
\\
& \left(\frac{\dis \langle \pi\rangle}{\dis \langle N_p\rangle} \right)_{p+p}~>~
\left(\frac{\dis \langle \pi\rangle}{\dis \langle N_p\rangle} \right)_{Pb+Pb}~.
\label{NpiNp1}
\end{eqnarray}
This happens because of the  fusion reactions
$\text{meson}+\text{baryon}\to \text{baryon}$  in dense baryon
system created in nucleus-nucleus collisions.
As seen from Fig.~\ref{fig:Npi}, the change of (\ref{NpiNp1}) to (\ref{NpiNp})
is indeed observed from the p+p and Pb+Pb
data, but at essentially larger collision energy $E_{\rm lab}\ge 40$A~GeV.

\section{Pion Spectra }
\label{sec-spec}

\subsection{Rapidity Spectra}
\label{rap}
The rapidity spectra $dN/dy$ of $\pi^-$ in inelastic p+p collisions \cite{NA61pp}
and central Pb+Pb collisions \cite{NA49-1,NA49-2} are presented
in Fig.~\ref{fig:dndycompare} for different collision energies.
The lines in this figure correspond to the results of the UrQMD simulations.
One can see that UrQMD underestimates the total yield of $\pi^-$ in p+p at all
energies except for the highest one,  $p_{\rm lab} = 158A$~GeV/c.
The deviations are bigger at the mid-rapidity.
This fact  was also noted in Ref.~\cite{Uzhinsky1}.
Our results for central Pb+Pb collisions are consistent with those published
earlier within the UrQMD-2.3 version~\cite{UrQMD2009NA49}.

\begin{figure}
\begin{center}
\begin{minipage}{.49\textwidth}
\centering
\includegraphics[width=\textwidth]{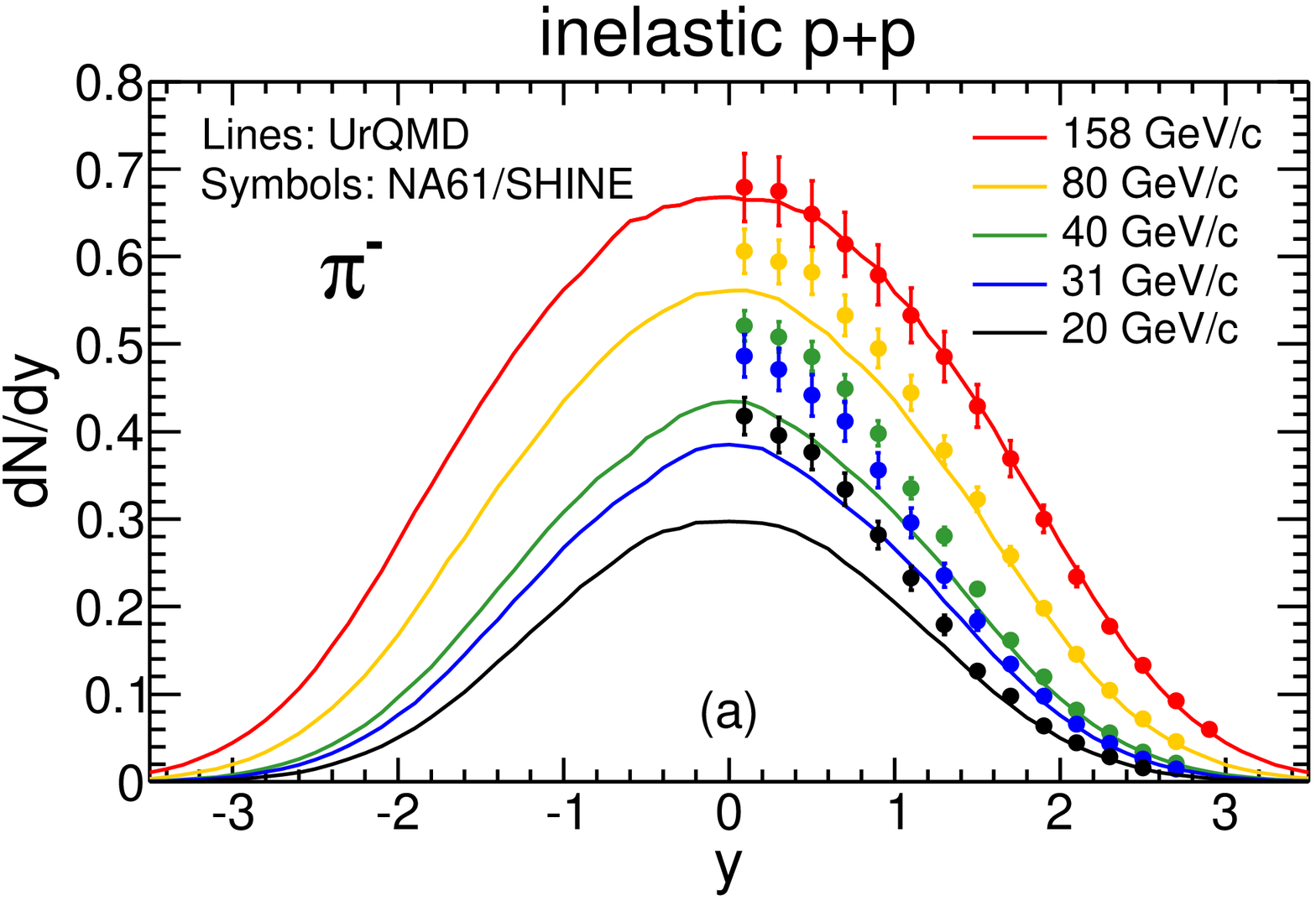}
\end{minipage}
\begin{minipage}{.49\textwidth}
\centering
\includegraphics[width=\textwidth]{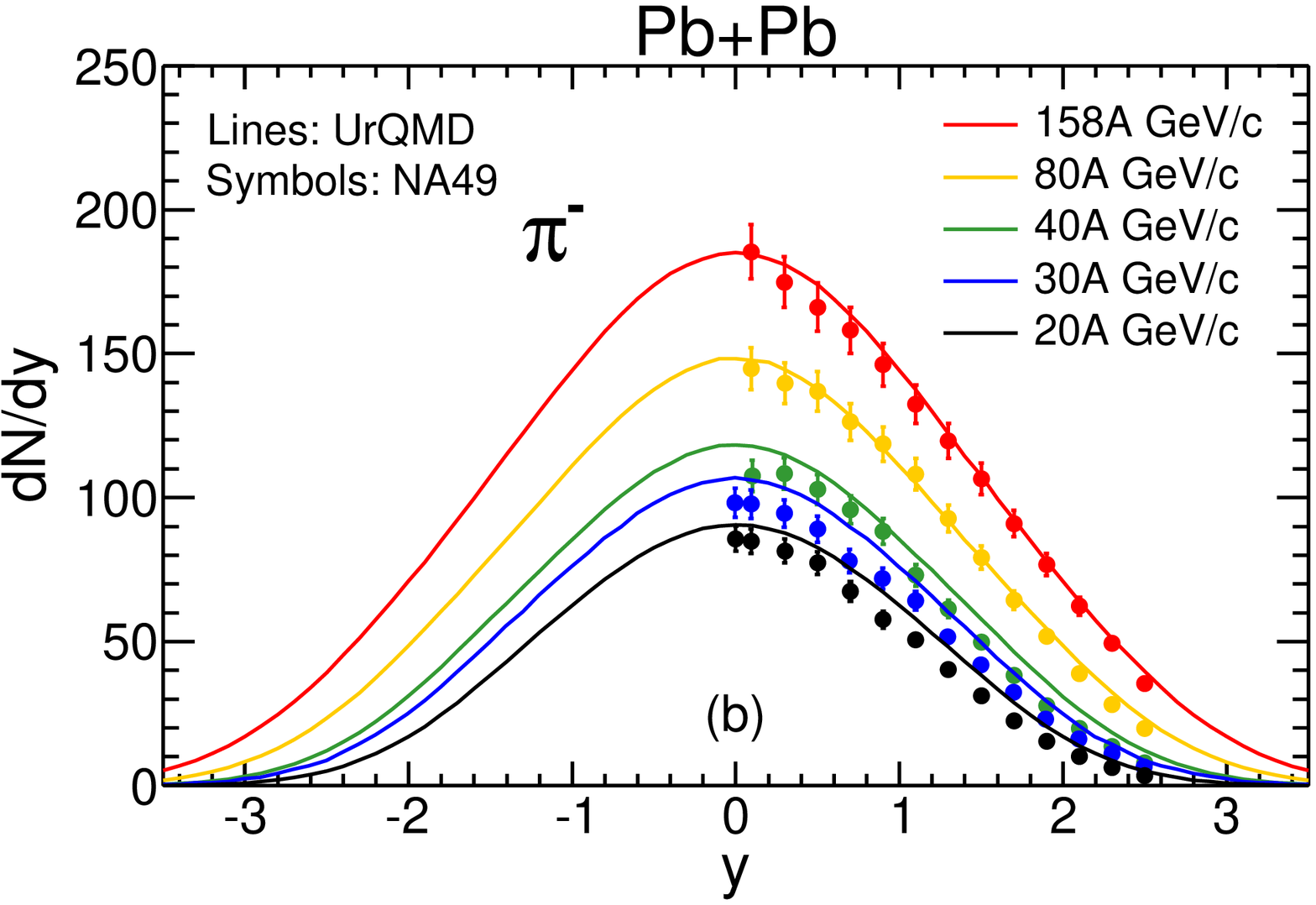}
\end{minipage}
\caption{(Color online) The rapidity distributions of $\pi^-$
in (a) inelastic p+p and (b) central Pb+Pb collisions. The symbols with error bars
correspond to the experimental data of NA61/SHINE and NA49 collaborations.
The solid lines show the results of the UrQMD calculations. }
\label{fig:dndycompare}
\end{center}
\end{figure}

As seen from Fig.~\ref{fig:dndycompare}b the UrQMD description
of the pion production appears to work much better in central Pb+Pb collisions.
This fact looks rather suspicious as the proton-proton results are the input
for the UrQMD cascade version description of nucleus-nucleus collisions. It means that
introducing the changes in a theoretical description of the p+p
results to achieve an agreement with the NA61/SHINE data \cite{NA61pp},
one will definitely  need to make
simultaneously additional changes to preserve the UrQMD agreement with the
nucleus-nucleus data.

\begin{figure}
\begin{center}
\begin{minipage}{.6\textwidth}
\centering
\includegraphics[width=\textwidth]{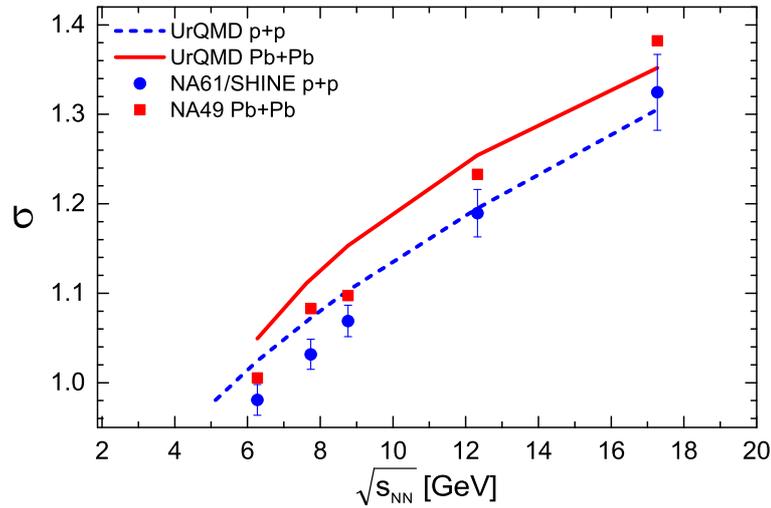}
\end{minipage}
\caption{(Color online) The energy dependence of the
rapidity width $\sigma$ for the $\pi^-$ spectra in inelastic p+p
and  central Pb+Pb collisions. The data of NA49 and NA61/SHINE
collaborations are depicted by symbols while the lines correspond to the UrQMD
calculations. }
\label{fig:RapidityWidthPi}
\end{center}
\end{figure}

The width of the rapidity distribution $dN/dy$ is an important feature
of the production process.
The width of the rapidity distribution
is defined as
\eq{\label{sigma1}
\sigma~ \equiv~ \sqrt{\langle y^2 \rangle~ -~ \langle y \rangle^2}~,
}
where ($k=1,2$)
\begin{equation}
\langle y^k \rangle~ = ~\frac{\int dy \, y^k \, (dN/dy)}{\int dy \, (dN/dy)}~.
\end{equation}
In the center of mass system, $\langle y \rangle = 0$ in a case of
the p+p collisions or collisions of identical nuclei.
The dependence of the
rapidity width $\sigma$ of the $\pi^-$ distribution $dN/dy$ on the collision energy
in inelastic  p+p and central Pb+Pb collisions is shown
in Fig.~\ref{fig:RapidityWidthPi}. One observes that
\eq{\label{sigma2}
\sigma({\rm Pb+Pb})~>~ \sigma({\rm p+p})~
}
for the whole SPS energy range. The UrQMD results are in a qualitative agreement
with experimental relation (\ref{sigma2}). However, the UrQMD values of $\sigma({\rm p+p})$
overestimate the experimental ones at the low SPS energies.
This is because of a deficiency
of pions in the mid-rapidity region in the UrQMD results seen in  Fig.~\ref{fig:dndycompare}a.

Note that the collision energy dependence of the rapidity distribution width
for different hadron species in Au+Au collisions was recently
studied within the UrQMD approach in Ref.~\cite{Dey2014}.

\subsection{Transverse Mass Distributions}
In Fig.~\ref{fig:transversecompare} the transverse mass distributions of
$\pi^-$ mesons are shown for inelastic p+p ({\it left}) and central Pb+Pb ({\it right})
collisions.
The symbols with error bars correspond to the experimental data of NA61/SHINE
\cite{NA61pp} and NA49 \cite{NA49-1,NA49-2} collaborations.
\begin{figure}
\begin{center}
\begin{minipage}{.49\textwidth}
\centering
\includegraphics[width=\textwidth]{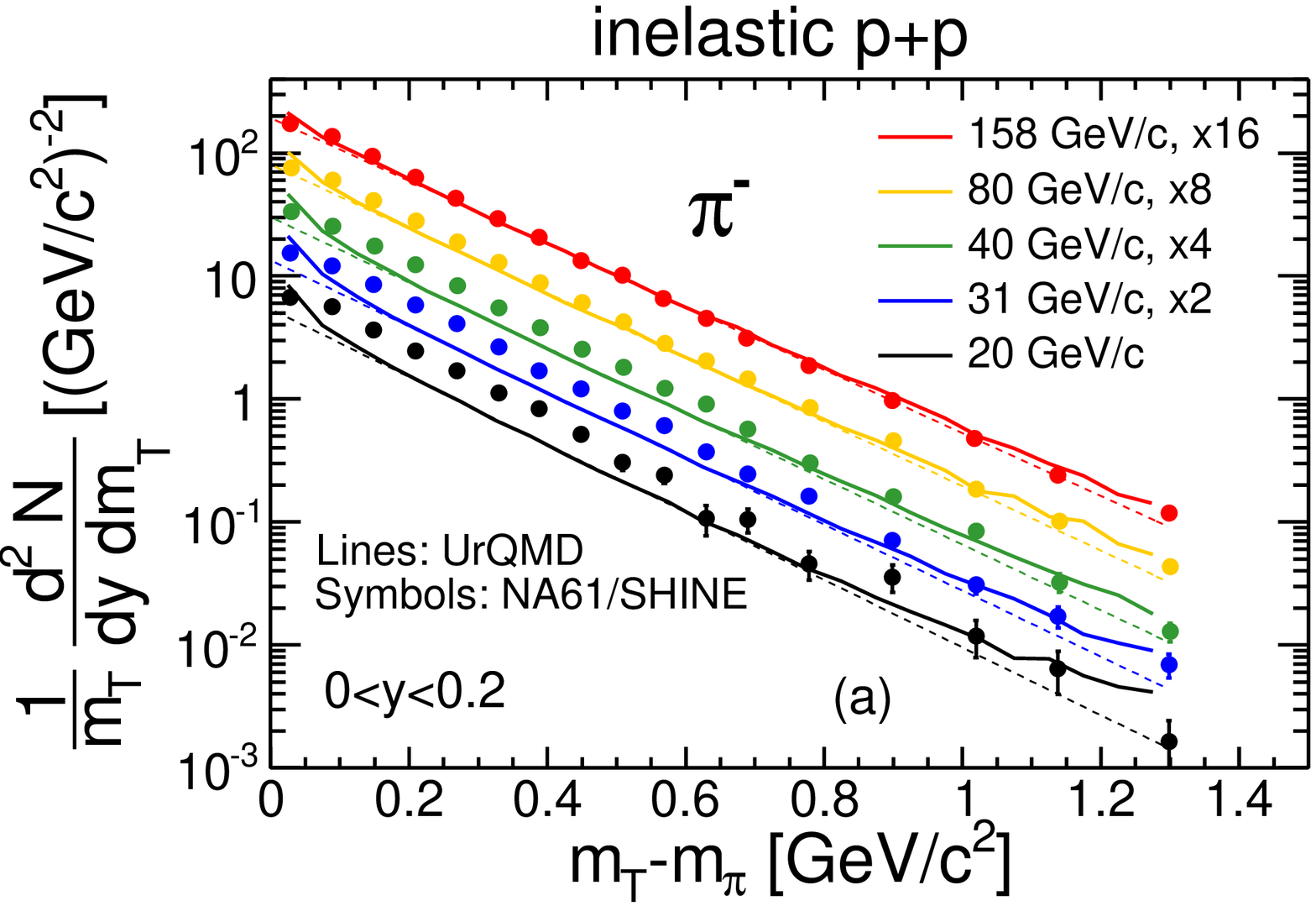}
\end{minipage}
\begin{minipage}{.49\textwidth}
\centering
\includegraphics[width=\textwidth]{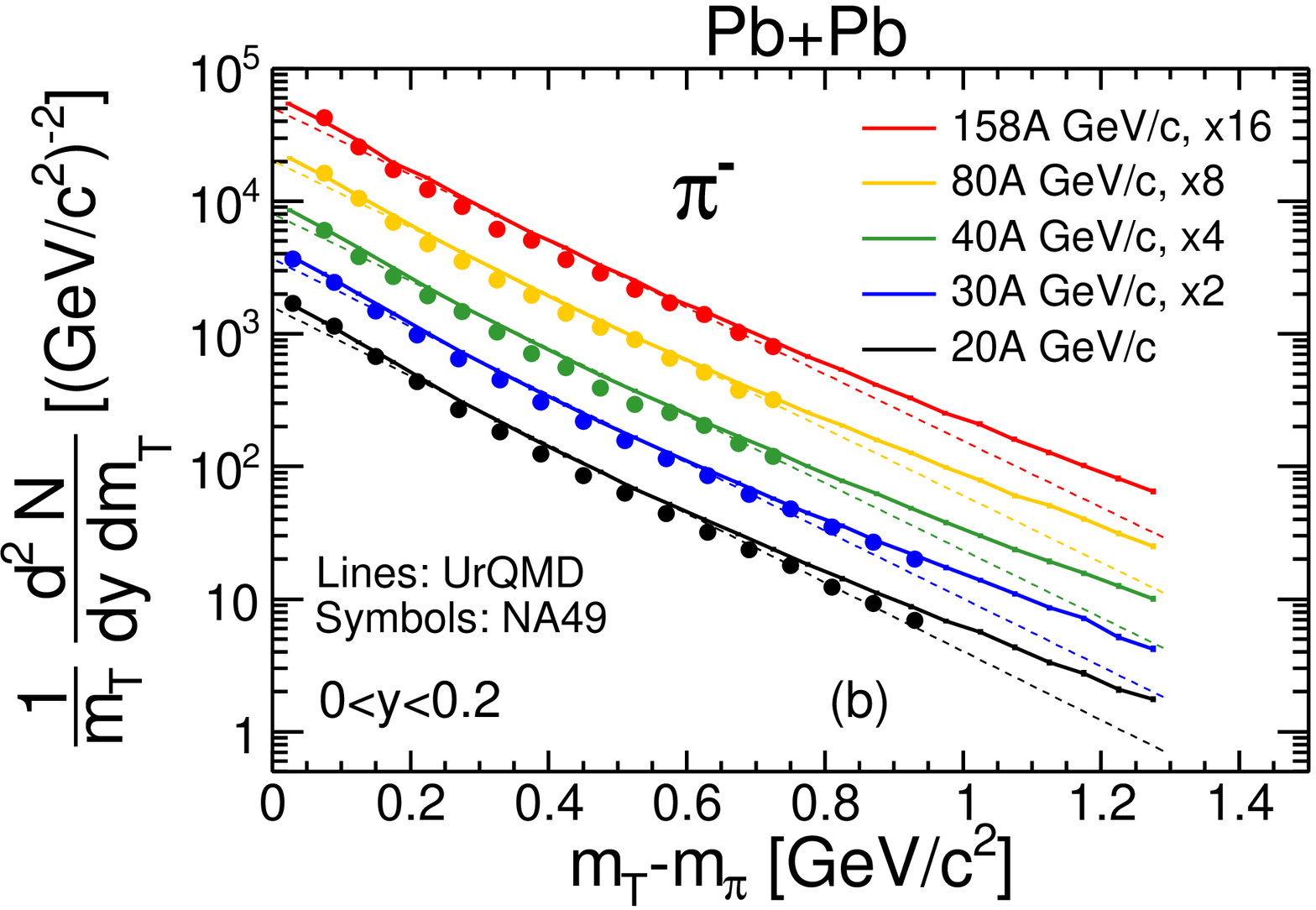}
\end{minipage}
\caption{(Color online) The transverse mass distributions of $\pi^-$
in (a) inelastic p+p and (b) central Pb+Pb collisions.
The solid lines correspond to the UrQMD calculations while the symbols with error bars
correspond to the experimental data of NA61/SHINE and NA49 collaborations.
The dotted lines correspond to the exponential fit (\ref{eq:dndmt}) of the UrQMD data
in the transverse mass interval $0.2<m_T-m_\pi<0.7$~GeV/c$^2$.}
\label{fig:transversecompare}
\end{center}
\end{figure}
The data correspond to the mid-rapidity spectra with $0<y<0.2$.
The solid lines correspond to the UrQMD calculations.
In the transverse mass region $0.2<m_T-m_\pi<0.7$~GeV/c$^2$,
the spectra for the both p+p and Pb+Pb reactions, and at all collision energies,
can be nicely fitted by a simple exponential function
\begin{equation}
\frac{dN}{m_Td m_T} ~=~ A\,\exp\left(-~\frac{m_T}{T}\right)~,
\label{eq:dndmt}
\end{equation}
where $A$ and $T$ are the fitting parameters, $m_T\equiv \sqrt{p_T^2+m_\pi^2}$
is the transverse mass with $p_T$ being the pion transverse momentum and
$m_\pi$ the pion mass.
The dependence of the inverse slope parameter $T$ of the $\pi^-$ transverse mass
spectra on the collision energy in p+p and Pb+Pb collisions is depicted
in Fig.~\ref{fig:TPi}a.
From this figures one observes that
\begin{equation}
T({\rm p+p})~<~ T({\rm Pb+Pb})~,
\label{T}
\end{equation}
for the whole SPS energy range.
The UrQMD results are in a qualitative agreement with experimental relation (\ref{T}).
However, UrQMD  overestimates systematically the experimental 
values of the inverse slope $T$ in p+p collisions at all SPS energies.

\begin{figure}
\begin{center}
\begin{minipage}{.49\textwidth}
\centering
\includegraphics[width=\textwidth]{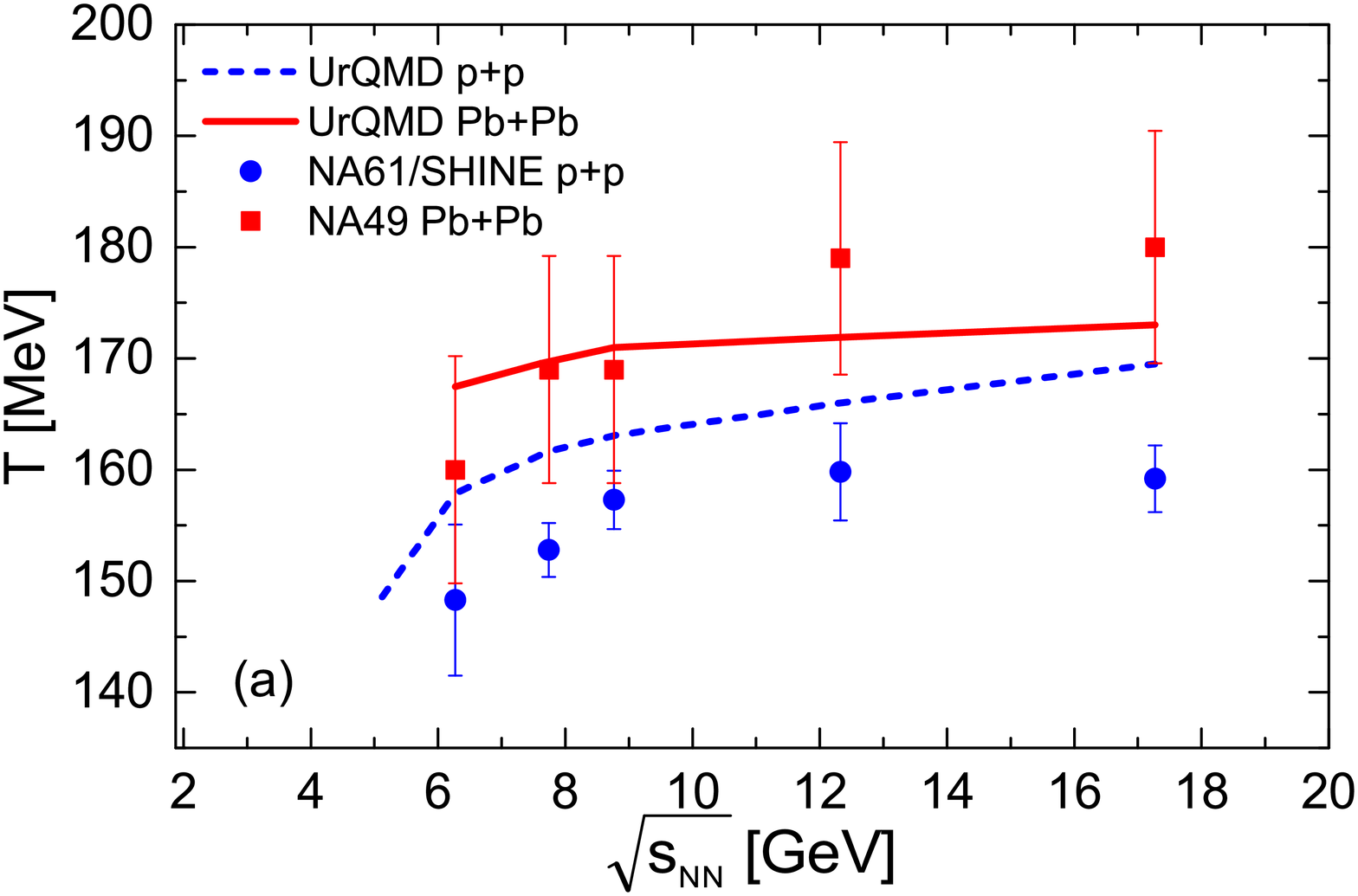}
\end{minipage}
\begin{minipage}{.49\textwidth}
\centering
\includegraphics[width=\textwidth]{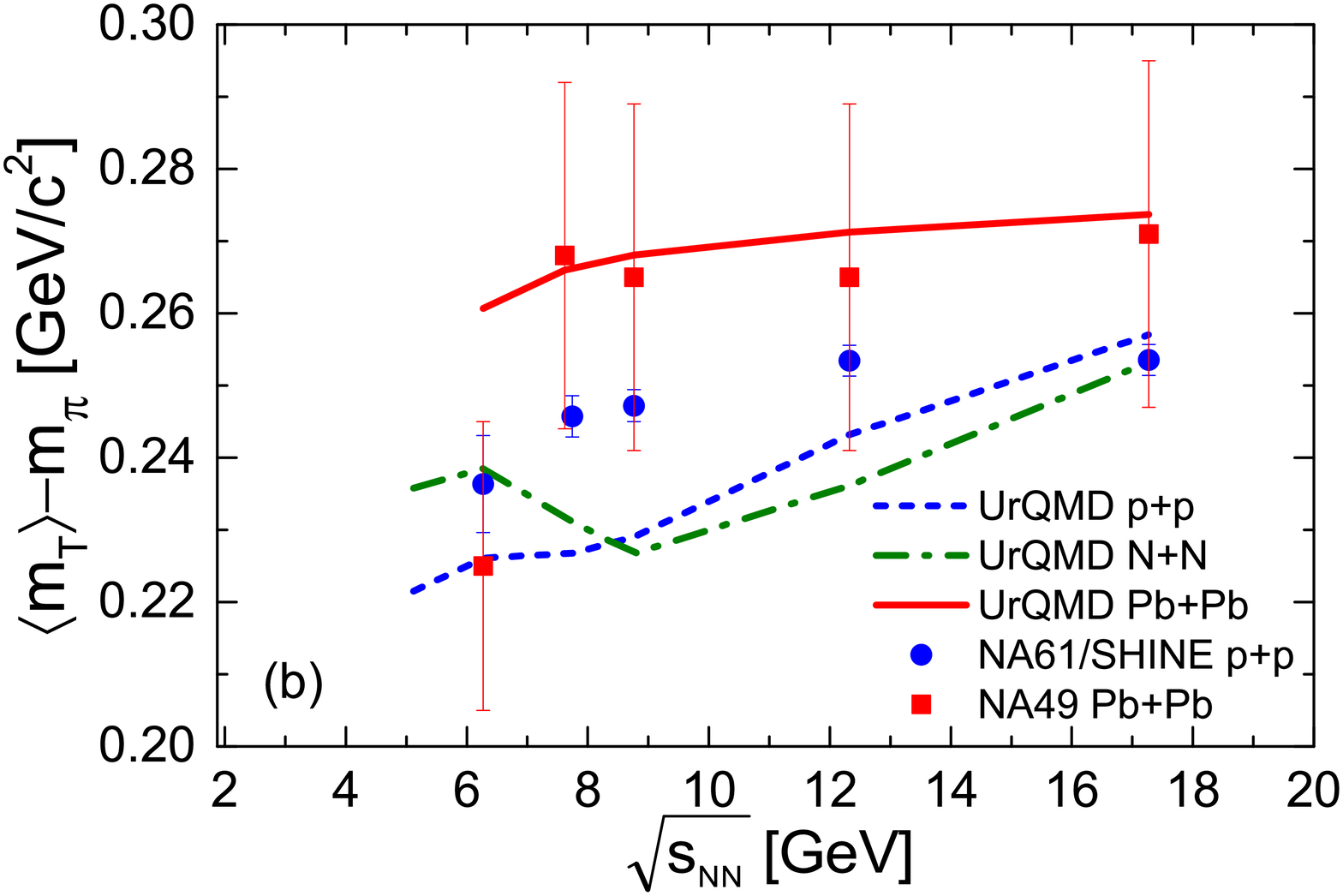}
\end{minipage}
\caption{(Color online) The energy dependence of (a) the inverse slope
parameter $T$ and (b) the mean transverse mass $\langle m_T\rangle$
of the $m_T$-spectra of $\pi^-$
at the mid-rapidity in inelastic p+p
and central Pb+Pb collisions. The lines correspond to the UrQMD
calculations while the experimental data of NA49 and NA61/SHINE
collaborations are depicted by the symbols with error bars.}
\label{fig:TPi}
\end{center}
\end{figure}
\begin{figure}
\begin{center}
\begin{minipage}{.65\textwidth}
\centering
\includegraphics[width=\textwidth]{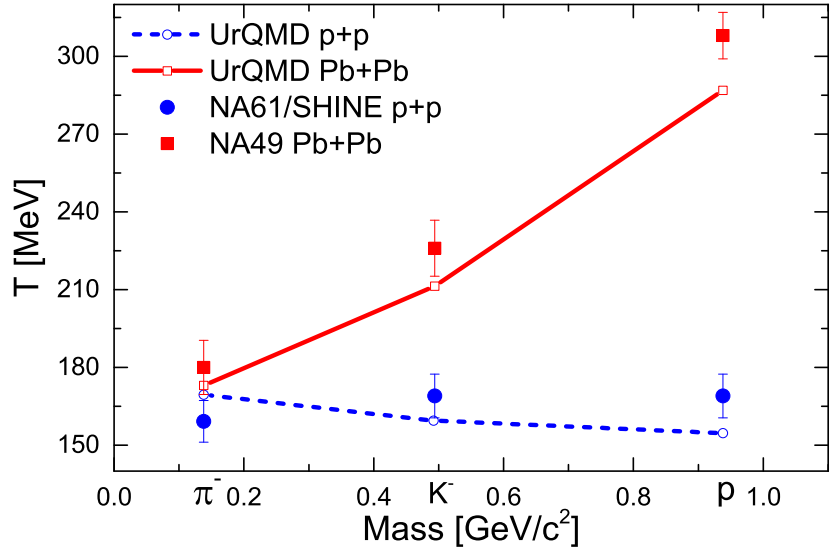}
\end{minipage}
\caption{(Color online) Hadron mass dependence of the inverse slope
parameter at mid-rapidity in inelastic p+p
and central Pb+Pb collisions at $p_{\rm lab}=158A$~GeV/c.
The open symbols correspond to the UrQMD
calculations while the data of NA49 \cite{NA49-1,NA49-3,NA49-2}
and preliminary results of NA61/SHINE \cite{NA61-T}
collaborations are depicted by the full symbols.}
\label{fig:TvsM}
\end{center}
\end{figure}
The inverse slope parameter of the $m_T$-spectra is dependent on the hadron mass.
For inelastic p+p and central Pb+Pb collisions at $p_{\rm lab}=158A$~GeV/c this dependence
is presented in Fig.~\ref{fig:TvsM}.
The NA49 data \cite{NA49-1,NA49-3,NA49-2} correspond to the $m_T$-spectra of $\pi^-$, $K^-$, and p
at the mid-rapidity:
$0<y<0.2$ for $\pi^-$,
$-0.1<y<0.1$ for $K^-$, and $-0.5<y<-0.1$ for $p$.
In central Pb+Pb collisions, one observes the following inequalities of the inverse slopes:
\begin{equation}
T_\pi~<~T_K~<~T_{\rm p}~,
\label{T1}
\end{equation}
i.e.,  rather strong  increase of the $T$ parameter with particle mass.
In the language of hydrodynamics this increase reflects  a presence
of the transverse collective flow in nucleus-nucleus collisions: at the same transverse
velocity of the hadronic fluid the particle with larger mass possesses  the larger transverse
momenta.

The preliminary results of NA61/SHINE \cite{NA61-T} correspond to
very different behavior
in inelastic p+p collisions:
\begin{equation}
T_\pi~\cong~T_K~\cong~T_{\rm p}~.
\label{T2}
\end{equation}
Instead of a strong increase of $T$ with hadron mass in Pb+Pb collisions,
the value of $T$  is approximately  constant for $\pi^-$, $K^-$, and p
in p+p collisions.

Next, we investigate a behavior of the mean transverse mass.
In Fig.~\ref{fig:TPi}b the mean transverse mass of $\pi^-$
at the mid-rapidity is shown as the
function of collision energy in p+p and Pb+Pb collisions.
Similar to Eq.~(\ref{NN}) one obtains for the mean transverse mass $\langle m_T\rangle$
of $\pi^-$ or $\pi^+$ in nucleon-nucleon collisions:
\begin{equation}
\langle m_T^{NN} \rangle~
= ~\alpha_{pp} \, \langle m_T^{pp} \rangle
~+~ \alpha_{pn} \, \langle m_T^{pn} \rangle
~ +~ \alpha_{nn} \, \langle m_T^{nn} \rangle\,,
\label{NNmT}
\end{equation}
In Fig.~\ref{fig:AvMtinelNN} the UrQMD results are presented for the mean
transverse mass $\langle m_T\rangle$ of $\pi^-$ ({\it left}) and
$\pi^+$ ({\it right}) at mid-rapidity ($0<y<0.2$) in inelastic p+p, p+n, and n+n collisions.
One observes a prominent decrease of $\langle m_T\rangle$ for $\pi^-$ in n+n
collisions in the range of laboratory energy between 20 and 40~GeV, and a
similar effect for $\pi^+$ in p+p collisions.
These effects are really important for any analysis of the data in both p+p and
A+A collisions.
\begin{figure}
\begin{minipage}{.48\textwidth}
\centering
\includegraphics[width=\textwidth]{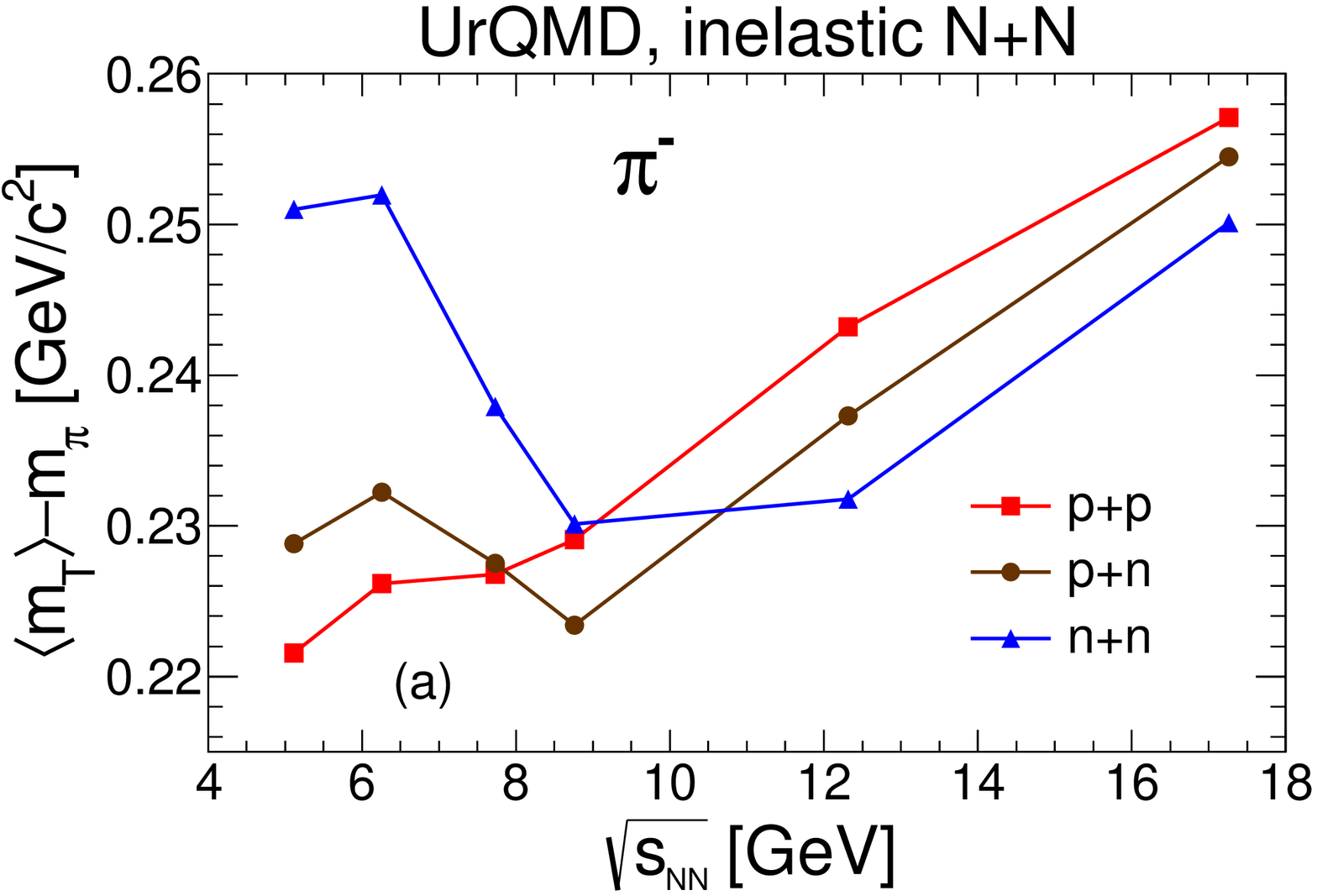}
\end{minipage}
\begin{minipage}{.48\textwidth}
\includegraphics[width=\textwidth]{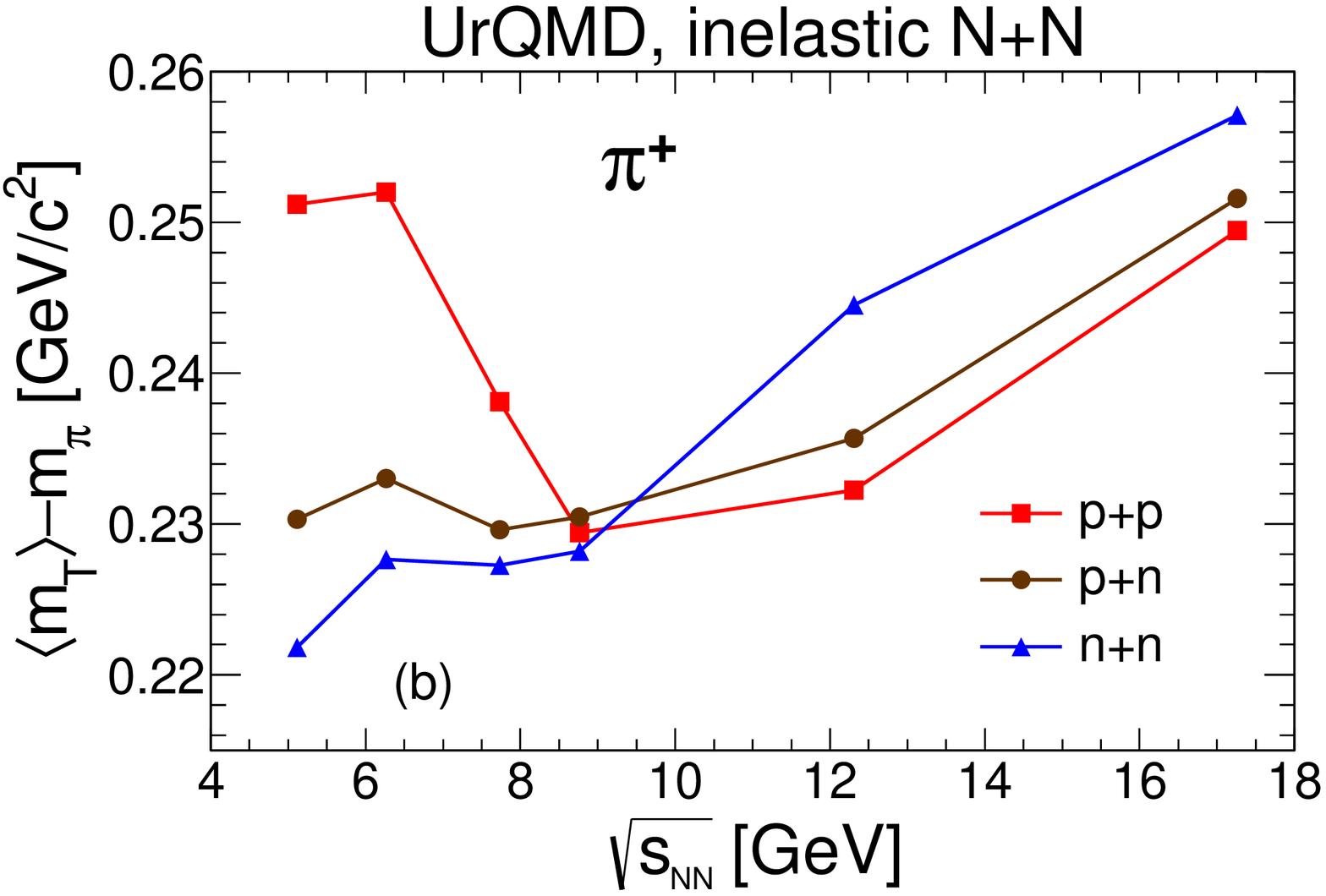}
\end{minipage}
\caption{(Color online) The points are the UrQMD results for
the mean transverse mass $\langle m_T\rangle$
of (a) $\pi^-$ and
(b) $\pi^+$ at mid-rapidity ($0<y<0.2$) in inelastic p+p, p+n and n+n.
The UrQMD data are depicted by the symbols, and the lines are drawn to guide the eye. }
\label{fig:AvMtinelNN}
\end{figure}

\begin{figure}
\begin{minipage}{.48\textwidth}
\centering
\includegraphics[width=\textwidth]{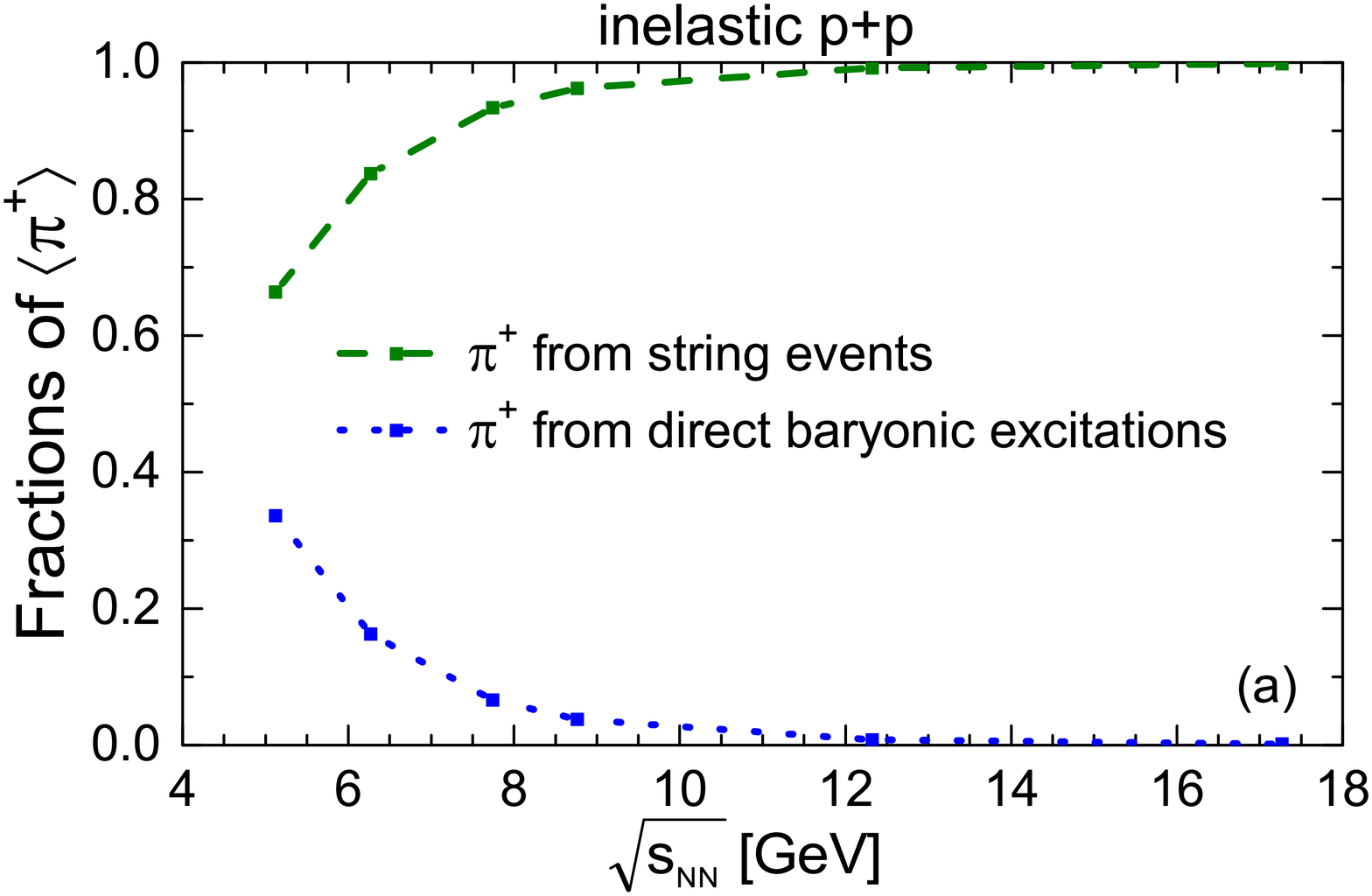}
\end{minipage}
\begin{minipage}{.48\textwidth}
\includegraphics[width=\textwidth]{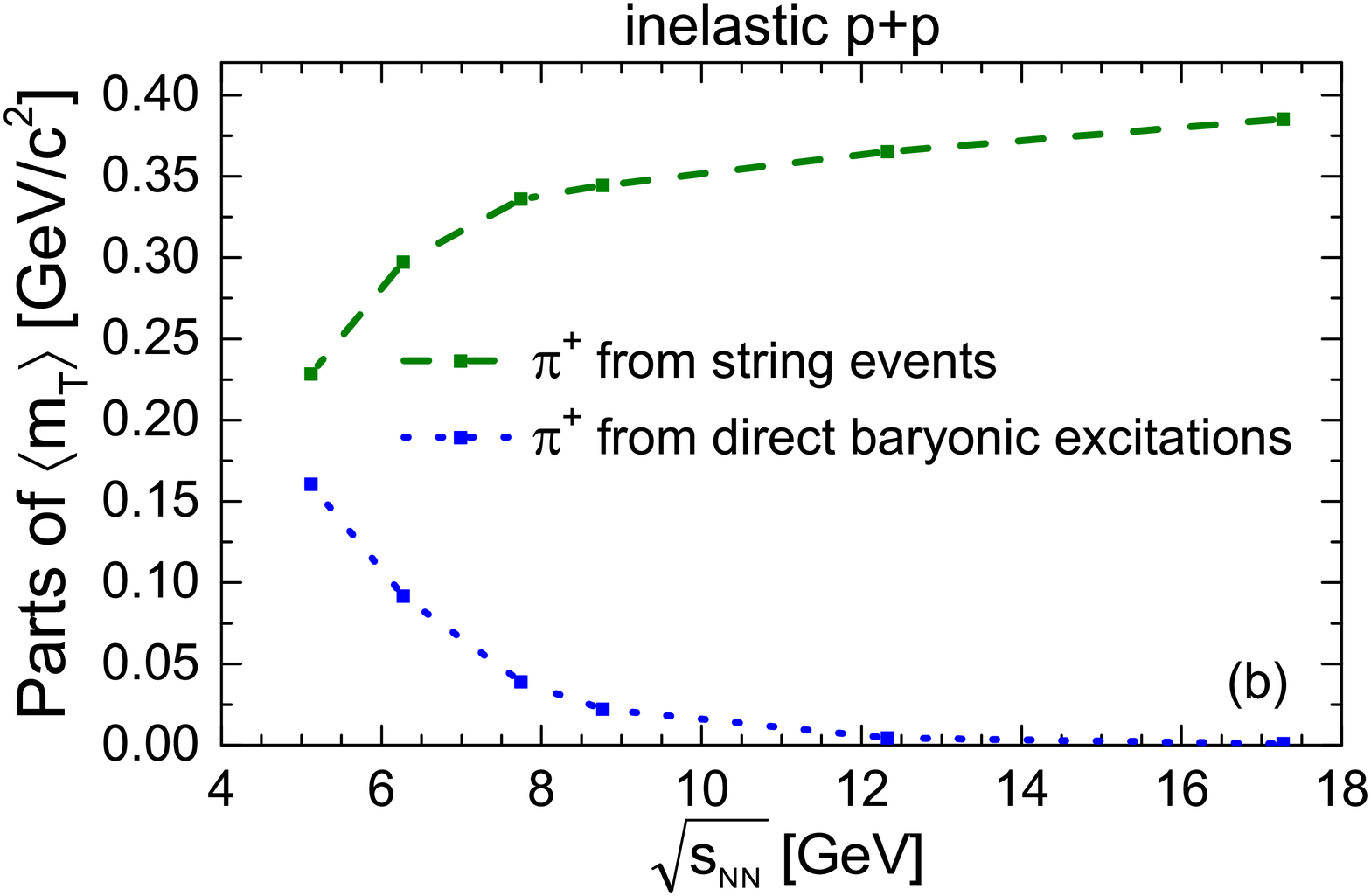}
\end{minipage}
\caption{(Color online) The UrQMD results for the $\pi^+$ production at the mid-rapidity
($0<y<0.2$) in inelastic p+p collisions. The contributions from the two sources of  $\pi^+$
production are shown:
1) the baryonic resonances produced according to Eq.~(\ref{pppi}) (the dotted lines);
2) the string produced according to Eq.~(\ref{string}) (the dashed lines).
(a) The fractions of the $\pi^+$ multiplicity $f_B$ and $f_s$ from the two sources.
(b) The parts of the $\langle m_T\rangle$ value $f_B\, \langle m_T\rangle_B$
and $f_s\, \langle m_T\rangle_s$ from the two sources.
}
\label{fig:Delta}
\end{figure}
A physical origin of the non-monotonous  energy dependence of $\langle m_T\rangle$
for $\pi^-$ in n+n collisions and for $\pi^+$ in p+p collisions is connected to the
presence of two different sources of pion production: excitation and decay of
the baryonic resonances $N^*$ and $\Delta$, and excitation and decay of the hadronic strings.
In p+p collisions the main inelastic reactions at small energies are the following:
\begin{eqnarray}
&& p+p\rightarrow p+\Delta^+~,~~~~~p+p\rightarrow n+\Delta^{++}~,~~~~~
p+p\rightarrow \Delta^++\Delta^+~,~~~~~ p+p\rightarrow \Delta^0+\Delta^{++}~,
\nonumber \\
&& p+p\rightarrow p+N^+~,~~~~~ p+p\rightarrow N^++\Delta^+~,~~~~~p+p\rightarrow N^0+\Delta^{++}~.
\label{pppi}
\end{eqnarray}
In the UrQMD simulations, the states $N^*$ with $m=1440,\ldots, 2250$~MeV and
$\Delta$ with $m=1232,\ldots,1950$~MeV are included \cite{UrQMD1998,UrQMD1999}.
The present version of the UrQMD model includes also
the supplementary baryonic  resonances $N^*$ and $\Delta$
with artificially large masses \cite{UrQMD:2008}.
Searching for charged pions in the final state, one observes 
a difference in the production of $\pi^-$ and $\pi^+$:
only $\Delta^0$ and $N^0$ may produce $\pi^-$, while other baryonic resonances,
$\Delta^+$ and $N^+$, may produce $\pi^+$,
and $\Delta^{++}$ has to produce  $\pi^+$, but they can not produce $\pi^-$.

Reactions (\ref{pppi}) give the dominant contribution to the p+p inelastic
cross section at small collision energies.
However, at $\sqrt{s_{NN}}\ge 4$~GeV  the excitation of strings,
\begin{equation}
p~+~p~\rightarrow ~{\rm string}~+~ {\rm string}~,
\label{string}
\end{equation}
opens the new channels of pion production. At $\sqrt{s_{NN}}\ge (6\div 8)$~GeV
the string production dominates in the UrQMD description of the inelastic p+p
cross section \cite{UrQMD1998}.

In Fig.~\ref{fig:Delta} we present two  contributions to the mean multiplicity
({\it left}) and to mean transverse mass $\langle m_T\rangle$ ({\it right})
for $\pi^+$ at the mid-rapidity, $0<y<0.2$, calculated in p+p inelastic collisions
within UrQMD simulations:~ 1) the excitation of baryonic resonances according to
Eq.~(\ref{pppi}) and their decays to $\pi^+$ plus nucleon;~ 2) the excitation of
strings according to Eq.~(\ref{string}) and their decays to $\pi^+$ plus anything.

The value of $\langle m_T\rangle$ can be presented as
\begin{equation}
\langle m_T\rangle~=~ f_B \, \langle m_T\rangle_B~+~f_s\, \langle m_T\rangle_s~.
\label{partial}
\end{equation}
In Eq.~(\ref{partial}), $f_B$ and $f_s$ correspond to the fractions of $\pi^+$
numbers from excited baryons (\ref{pppi}) and strings (\ref{string}),
respectively, whereas $\langle m_T\rangle_B$ and $ \langle m_T\rangle_s$ are the
corresponding values of the mean transverse mass from these two mechanisms.
From Fig.~\ref{fig:Delta}a one observes that $f_s>f_B$ at all SPS
energies, and  $f_s$ becomes much larger than $f_B$ with increasing energy.
However,
at small SPS energies this inequality is partially compensated by the opposite one,
$\langle m_T\rangle_B > \langle m_T\rangle_s $. The partial contributions
to $\langle m_T\rangle$ coming from baryonic and string decays defined by Eq.~(\ref{partial})
are shown in Fig.~\ref{fig:Delta}b.
The interplay of these two contributions leads to the non-monotonous behavior
of their sum $\langle m_T\rangle$ for $\pi^+$  in the region $E_{\rm lab}=20-40$~GeV
seen in Fig.~\ref{fig:AvMtinelNN}b.

\begin{figure}
\begin{minipage}{.48\textwidth}
\centering
\includegraphics[width=\textwidth]{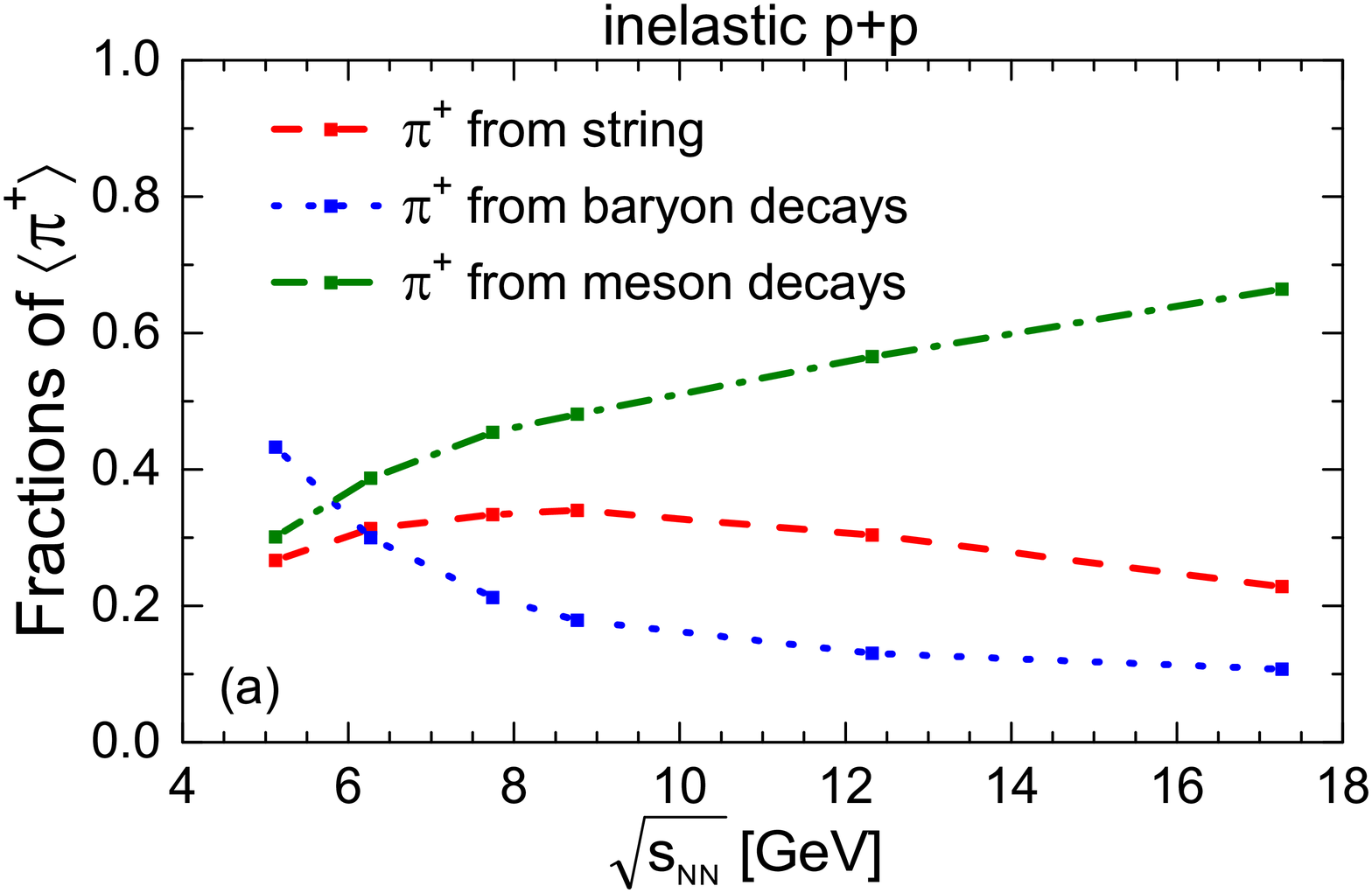}
\end{minipage}
\begin{minipage}{.48\textwidth}
\includegraphics[width=\textwidth]{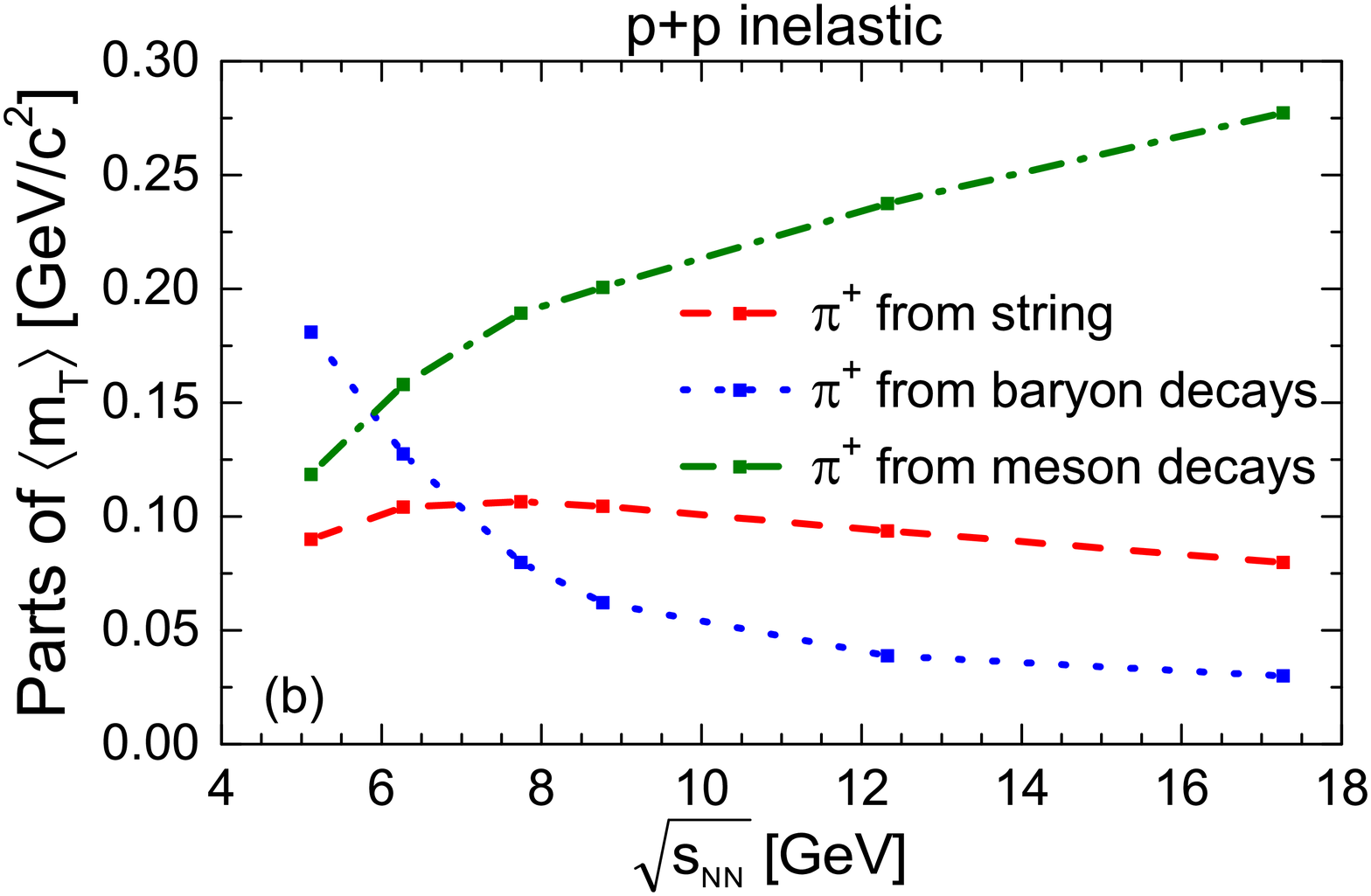}
\end{minipage}
\caption{(Color online)
The UrQMD results for the $\pi^+$ production at the mid-rapidity
($0<y<0.2$) in inelastic p+p collisions. The contributions from three sources of the $\pi^+$
production are shown:
1) all baryonic resonances (the dotted lines);
2) all mesonic resonances (the dashed-dotted lines);
3) the `direct' pions from the string decays (the dashed lines).
(a) The fractions of the $\pi^+$ multiplicity from the three sources. (b) The partial contributions to the $\langle m_T\rangle$ value from the three sources.
}
\label{fig:3sources}
\end{figure}
It seems to be also instructive to distinguish three different sources
of pion production: all baryonic resonances (including the direct ones produced according
to Eq.~(\ref{pppi}) and those appeared from  decays of the strings), all mesonic resonances
(appeared from the string fragmentations and from decays of the highly excited baryonic resonances),
and the `direct' pions from string decays (without intermediate resonance states).
In Fig.~\ref{fig:3sources} these three contributions calculated within UrQMD
in p+p inelastic collisions are presented.
The {\it left} panel (a) shows  the fractions of the mean multiplicity  and  the
{\it right} panel (b) the partial contributions to $\langle m_T\rangle$ value for
$\pi^+$  at the mid-rapidity, $0<y<0.2$.
The decrease of $\langle m_T\rangle$ of $\pi^+$ with collision energy seen  in
Fig.~\ref{fig:AvMtinelNN}b appears again as a result of the
interplay of these three contributions: a strong decrease with collision energy
of the baryonic contribution, an
increase of the contribution from the mesonic resonances,
and approximately independent of  collision energy
the contribution of the `direct' pions from the strings.

Note that our observation of the non-monotonous behavior of $\langle m_T\rangle$
for $\pi^+$ with collision energy at $20<E_{{\rm lab}}<40$~GeV in inelastic p+p collisions
shown in Fig.~\ref{fig:AvMtinelNN}b
is based on the UrQMD  results.
On the other hand, as it was discussed above, the UrQMD results
underestimate the yield
of $\pi^-$ in p+p at  $p_{{\rm lab}} \le 80$~GeV/$c$. Therefore,
some improvements of the UrQMD model are desirable.
As it was mentioned, the present version of the UrQMD code includes
many supplementary baryonic  resonances $N^*$ and $\Delta$
in Eq.~(\ref{pppi}) with artificially large masses \cite{UrQMD:2008}.
These resonances influence the
UrQMD results for the $\langle m_T\rangle$ of pions at SPS energies.
However, a necessity of their presence deserves further studies.
It was shown recently in Ref.~\cite{Uzhinsky1} that an agreement
between the UrQMD simulations and the
data at the SPS energies
can be improved by suppressing the probabilities
of binary inelastic reactions (\ref{pppi}). Such a modification will lead
automatically to the enhanced
contribution from the
string excitation processes (\ref{string}). The modified UrQMD code
will inevitably change an interplay of the contributions from the  baryonic  resonances
and excited strings.

Note also that the observed non-monotonous behavior of $\langle m_T\rangle$ is  specific
for the pions accepted at the mid-rapidity. The UrQMD calculations for all secondary
pions presented in Fig.~\ref{fig:all-y} do not show any non-monotonous energy dependence.
This is in an agreement with the compilation of old p+p data in Ref.~\cite{compil}.
The dependence of the non-monotonous behavior of $\langle m_T\rangle$ on
the size of the rapidity window for the accepted $\pi^+$ and other details
will be studied further  in Ref.~\cite{VAG}.
\begin{figure}
\begin{minipage}{.48\textwidth}
\centering
\includegraphics[width=\textwidth]{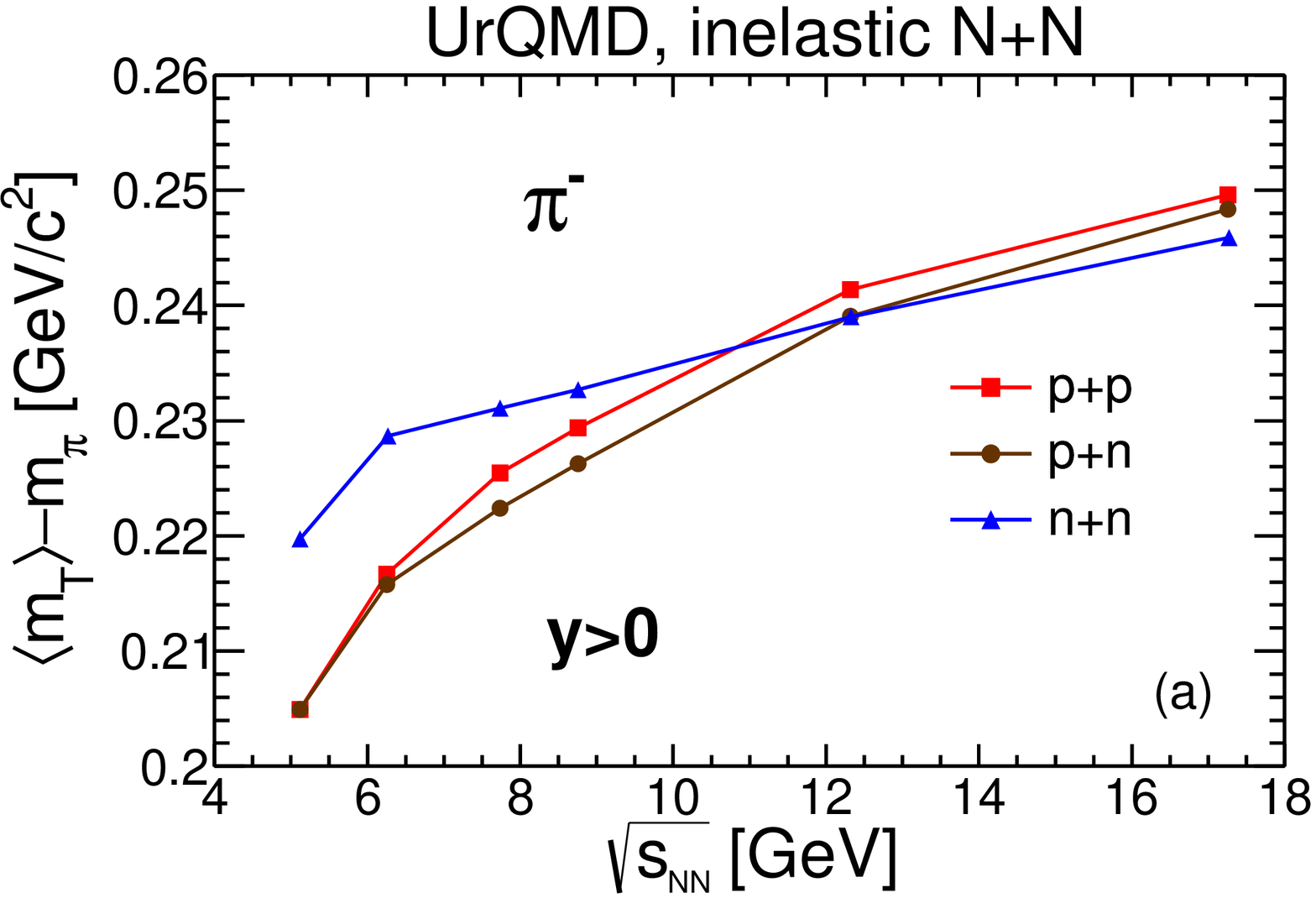}
\end{minipage}
\begin{minipage}{.48\textwidth}
\includegraphics[width=\textwidth]{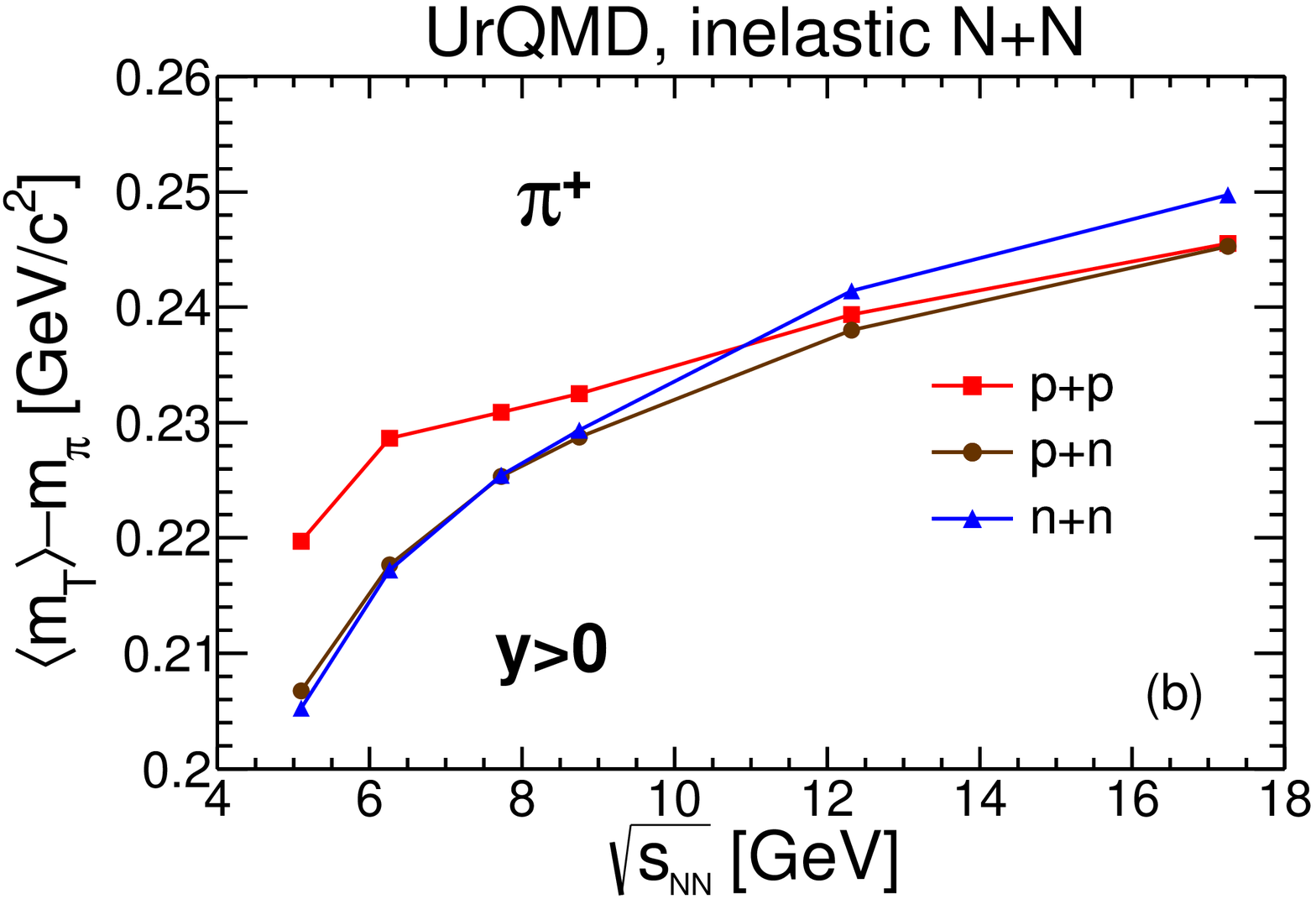}
\end{minipage}
\caption{(Color online) The points are the UrQMD results for
the mean transverse mass $\langle m_T\rangle$
of (a) $\pi^-$ and
(b) $\pi^+$ at all rapidities $y>0$ in inelastic $p+p$, $p+n$ and $n+n$.
The UrQMD data are depicted by the symbols, and the lines are drawn to guide the eye. }
\label{fig:all-y}
\end{figure}
In any case, our UrQMD prediction for the $\langle m_T\rangle$ drop
for $\pi^+$ at the mid-rapidity at $20<p_{\rm lab}<40$~GeV/c in inelastic p+p collisions
will be checked soon
experimentally by the NA61/SHINE Collaboration.

\section{Summary}
\label{sum}

In the present paper, the UrQMD simulations are
exploited
for a theoretical modeling of the
collision energy and system-size
effects for
the $\pi^-$ spectra in inelastic p+p collisions  measured  recently
by the NA61/SHINE \cite{NA61pp}  and in central Pb+Pb collisions
\cite{NA49-1,NA49-3,NA49-2}.

The UrQMD simulations demonstrate that at SPS energies ($E_{\rm lab}\geq 10A$~GeV) each participant
nucleon in central nucleus-nucleus collisions transforms more energy
to the production of new particles than that in p+p inelastic
collisions  at the same initial energy per nucleon.
At small collision energy $E_{\rm lab}\le 10A$~GeV the UrQMD simulations show
however the opposite behavior: the energy loss per participant nucleon
$E_{\rm loss}$ becomes larger in inelastic p+p collisions.
This UrQMD result is correlated to the inequality between the  mean pion
multiplicities per participant nucleon, $\left(\langle \pi\rangle/
\langle N_p\rangle\right)_{p+p}$ and $\left(\langle \pi\rangle/
\langle N_p\rangle\right)_{Pb+Pb}$.
A change of the sign in the difference of the pion multiplicities per nucleon
in p+p and Pb+Pb collisions is indeed observed from the
data, but at essentially larger collision energy $E_{\rm lab}\cong 40A$~GeV.

The UrQMD results underestimate the total yield of $\pi^-$ at the mid-rapidity
in inelastic p+p collisions  at all SPS energies except for the highest one,
$p_{\rm lab} = 158A$~GeV/c.
The UrQMD description appears to work better in central Pb+Pb collisions.
This observation deserves a further analysis.
The p+p results are the input for the UrQMD cascade version description of
nucleus-nucleus collisions.
Thus, any changes of this input,
introduced to fit the new p+p data at SPS energies,
may destroy an existing agreement with the Pb+Pb data.

The width of the rapidity spectra and the inverse slope of the transverse momentum
spectra or the mean transverse mass
were investigated on the base of the UrQMD simulations and compared to the
data.
Both the rapidity width and inverse slope for $\pi^-$ spectra in Pb+Pb
collisions are larger than those in p+p collisions, and for both reactions they
increase monotonously with the collision energy.
The UrQMD results are in a qualitative agreement with
experimental observations.

An important aspect of our analysis is a proper treatment of the isospin effects.
They appear to be important for a comparison of the pion production data in
proton-proton and nucleus-nucleus collisions.
We demonstrate that different results for the pion production in p+p, p+n, and
n+n collisions are important not only for $\pi^-$ and $\pi^+$ multiplicities
but, even more, for their $m_T$-spectra. 
Large difference of about $20-30$~MeV/c$^2$ in $\langle m_T\rangle$ values for
secondary $\pi^-$ in p+p and n+n inelastic collisions (and the same value with the opposite sign for
$\pi^+$) is predicted by the UrQMD
simulations at low SPS energies,
in the energy range corresponding to the future CBM experiment at FAIR.
This is
important
for the analysis of the $\langle m_T\rangle$ dependence on the collision energy
in nucleus-nucleus collisions.

The UrQMD simulations predict a non-monotonous behavior of
$\langle m_T\rangle$ with collision energy for $\pi^+$  in inelastic p+p reactions:
the mean transverse mass of positively charged pions accepted at mid-rapidity
decreases notably with the collision energy inside the
region of $p_{\rm lab}=20\div 40$~GeV/c.
This our prediction of rather unexpected strong drop of $\langle m_T\rangle$
for $\pi^+$ in p+p collisions will be checked soon by the NA61/SHINE Collaboration.

\begin{acknowledgments}
%
We would like to thank Elena Bratkovskaya,
Ivan Kisel,
Tobiasz Czopowicz, Marek Ga\'zdzicki, Pasi Huovinen, Szymon Pulawski, and Yurij Sinyukov for fruitful discussions and comments.
The work of D.V.A. and M.I.G. was supported  by the
Program of Fundamental Research of the Department of Physics and
Astronomy of NAS and by the State Agency of Science, Innovations and
Informatization of Ukraine contract F58/384-2013.
\end{acknowledgments}


\end{document}